\providecommand{\reserveinserts}[1]{}
\documentclass[AMA,STIX1COL]{WileyNJD-v2}
\usepackage{dsfont}

\usepackage[utf8]{inputenc}
\usepackage{amsthm,amsmath}
\usepackage{bbm}
\usepackage{cancel}
\usepackage{hyperref}
\usepackage{upgreek}
\DeclareMathOperator{\logit}{logit}
\DeclareMathOperator{\expit}{expit}
\usepackage{comment} 
\newcommand{\ind}{\perp\!\!\!\!\perp} 

\usepackage{setspace}
\articletype{}%

\received{26 April 2016}
\revised{6 June 2016}
\accepted{6 June 2016}

\begin{document}
\onehalfspacing

\title{Doubly robust augmented weighting estimators for the analysis of externally controlled single-arm trials and unanchored indirect treatment comparisons}

\author[1,2]{Harlan Campbell}

\author[3,4]{Antonio Remiro-Az\'ocar}

\authormark{CAMPBELL \& REMIRO-AZ\'OCAR}

\address[1]{\orgdiv{Evidence Synthesis and Decision Modeling}, \orgname{Precision AQ}, \orgaddress{\state{British Columbia}, \country{Canada}}}

\address[2]{\orgdiv{Department of Statistics}, \orgname{University of British Columbia}, \orgaddress{\state{British Columbia}, \country{Canada}}}

\address[3]{\orgdiv{External Collaboration and Experimentation}, \orgname{Novo Nordisk Pharma}, \orgaddress{\state{Madrid}, \country{Spain}}}

\address[4]{\orgdiv{Department of Statistical Science}, \orgname{UCL}, \orgaddress{\state{London}, \country{United Kingdom}}}

\corres{} 


\abstract{Externally controlled single-arm trials are critical to assess treatment efficacy across therapeutic indications for which randomized controlled trials are not feasible. A closely-related research design, the unanchored indirect treatment comparison, is often required for disconnected treatment networks in health technology assessment. We present a unified causal inference framework for both research designs. We develop an estimator that augments a popular weighting approach based on entropy balancing -- matching-adjusted indirect comparison (MAIC) -- by fitting a model for the conditional outcome expectation. The predictions of the outcome model are combined with the entropy balancing MAIC weights. While the standard MAIC estimator is singly robust where the outcome model is non-linear, our augmented MAIC approach is doubly robust, providing increased robustness against model misspecification. This is demonstrated in a simulation study with binary outcomes and a logistic outcome model, where the augmented estimator demonstrates its doubly robust property, while exhibiting higher precision than all non-augmented weighting estimators and near-identical precision to G-computation. We describe the extension of our estimator to the setting with unavailable individual participant data for the external control, illustrating it through an applied example. Our findings reinforce the understanding that entropy balancing-based approaches have desirable properties compared to standard ``modeling'' approaches to weighting, but should be augmented to improve protection against bias and guarantee double robustness.}

\keywords{Single-arm trial, indirect treatment comparison, external control, covariate adjustment, evidence synthesis, data fusion}

\maketitle
\thispagestyle{empty}

\clearpage

\renewcommand{\thefootnote}{\alph{footnote}}

\section*{Highlights}
\paragraph{What is already known?}

\begin{itemize}
\item Externally controlled single-arm trials are important for assessing treatment efficacy when randomized controlled trials are not feasible. 
\item  Unanchored indirect treatment comparisons, often using the entropy balancing or matching-adjusted indirect comparison (MAIC) approach, can provide treatment effect estimates for disconnected networks in health technology assessment.
\end{itemize}

\paragraph{What is new?}

\begin{itemize}
\item A unified causal inference framework for both externally controlled single-arm trials and unanchored indirect treatment comparisons is presented. 
\item A doubly robust estimator that augments the MAIC approach by combining outcome model predictions with entropy balancing weights is proposed. 
\item The proposed estimator provides increased protection against model misspecification, as demonstrated in a simulation study with binary outcomes. 
\item The proposed estimator is extended to the setting with unavailable individual participant data for the external control and illustrated through an applied example. 
\end{itemize}

\paragraph{Potential impact for RSM readers outside the authors’ field}

\begin{itemize}
\item Doubly robust augmented weighting estimators, particularly using entropy balancing weights, offer increased robustness to model misspecification than conventional non-augmented approaches for confounding adjustment. 
\item Entropy balancing-based methods have desirable properties compared to standard ``modeling'' approaches to weighting but should be augmented to improve protection against bias and guarantee double robustness. 
\end{itemize}

\section{Background}\label{sec1}

In pharmaceutical research, randomized controlled trials (RCTs) are the gold standard for evaluating treatment efficacy and safety due to their high internal validity. Random allocation minimizes confounding by balancing, in expectation, prognostic factors across treatment arms. Nevertheless, RCTs are not always feasible, for instance:
\begin{itemize}
\item Where recruitment is impractical due to small populations, e.g., rare diseases or biomarker-specific precision oncology;\cite{jahanshahi2021use}
\item For life-threatening conditions with high unmet need and inadequate standard of care, e.g., ``last-line of therapy'' indications in late-stage hematological and solid tumor oncology;\cite{mishra2022external}
\item Where placebo controls are unethical, e.g., withholding a therapy with proven efficacy in adults from a control group of children in pediatric trials.\cite{horton2021real} 
\end{itemize}

Regulatory agencies such as the Food and Drug Administration (FDA) and the European Medicines Agency (EMA) emphasize that RCTs provide the highest evidentiary standard.\cite{beaulieu2020examining, flynn2022marketing} However, regulators recognize that alternative designs may be required in special circumstances. One such design is the externally controlled single-arm trial (SAT), where the control group is fully derived from external data such as prior clinical trials or secondary real-world data (RWD) sources.\cite{mishra2022external, zou2024next, curtis2023regulatory, gray2020framework} Marketing authorization applications featuring externally controlled SATs continue to rise, especially under accelerated approval pathways.\cite{gray2020framework, sola2023effectively} In 2023 the FDA issued draft guidance for externally controlled trials,\cite{FDA2023guidance} and in 2024 the EMA finalized a reflection paper on SATs.\cite{EMA2024reflection} Approvals have been granted based on such designs, particularly for conditions with predictable natural history, precisely measurable endpoints, and anticipated large effect sizes.\cite{jahanshahi2021use, mishra2022external, gray2020framework, sola2023effectively, bakker2023contribution, goring2019characteristics}   

For health technology assessment (HTA), head-to-head RCT evidence also remains the gold standard.\cite{NICE2022manual, CADTH2024procedures, IQWiG2023methods} Nevertheless, as regulators increasingly approve pharmaceuticals based on externally controlled SATs, payer reliance on such designs has grown.\cite{sola2023effectively, patel2021use} HTA bodies are developing recommendations for externally controlled SATs,\cite{NICE2022framework, IQWiG2020guidance} with acceptability influenced by unmet need and disease rarity.\cite{sola2023effectively, patel2021use, jaksa2022comparison} HTA also requires comparing new technologies against all existing alternatives.\cite{paul2001fourth} The scope of assessments often depends on the policy question rather than available data\cite{EUnetHTA2021joint} and a single RCT cannot typically include all comparators desired for HTA, given the multiplicity of stakeholders and variations in clinical practice across jurisdictions.\cite{vreman2020differences} In the absence of direct RCT comparisons versus all
candidate comparators, indirect treatment comparisons (ITCs) are required.\cite{sutton2008use} 

HTA decision-makers prefer \textit{anchored} ITCs of randomized trials,\cite{phillippo2016nice, EUnetHTA2022methods, EUnetHTA2022practical} which respect randomization by using a common control arm to contrast relative treatment effects.\cite{phillippo2018methods} However, compatible control arms with which to ``anchor'' the analysis are not always available, especially in rapidly evolving areas with multiple novel treatments and no single accepted standard of care.\cite{goring2016disconnected, stevens2018review} In these scenarios, \textit{unanchored} ITCs based on disconnected networks may be required,\cite{goring2016disconnected, stevens2018review} and recent reviews determined that unanchored ITCs are in fact more common than anchored ITCs.\cite{serret2023methodological, truong2023population} Unanchored ITCs contrast mean treatment-specific absolute outcomes across studies, relying on more restrictive assumptions than anchored ITCs.\cite{phillippo2016nice, phillippo2018methods} In essence, they are externally controlled SATs where the external control is often a competitor's historical trial with limited data access: individual participant data (IPD) are available for the SAT, but only published aggregate-level data (AD) for the external control.\cite{phillippo2016nice, phillippo2018methods}

The absence of randomization compromises the validity of externally controlled SATs. Various statistical methods have been proposed to adjust for imbalances in baseline covariates \cite{lambert2023enriching} and these can potentially mitigate confounding bias and account for the additional variability induced by covariate differences.  The most widely-used methods are propensity score-based weighting approaches, typically using logistic regression estimated via maximum-likelihood.\cite{ren2023comparing, loiseau2022external, lunceford2004stratification, austin2016variance} For unanchored ITCs, matching-adjusted indirect comparison (MAIC) based on entropy balancing\cite{hainmueller2012entropy} is more popular.\cite{signorovitch2010comparative, cheng2020statistical, josey2021transporting} MAIC views covariate balance as a convex optimization problem, estimating weights that directly enforce balance without explicitly modeling the conditional probability of SAT participation.  MAIC is attractive for ITCs due to its applicability in IPD-AD situations and is thought to be more stable, precise, and robust to model misspecification than the standard propensity score-based weighting approaches, even in ``IPD-IPD'' scenarios.\cite{josey2021transporting, amusa2019examination, cheng2023double, zhao2017entropy} 

So-called ``G-computation'' or ``model-based standardization'' methods have also been developed for the IPD-IPD\cite{loiseau2022external, wang2017g, robins1986new, keil2014parametric} and IPD-AD settings.\cite{remiro2022parametric} For the latter, they are also referred to as ``simulated treatment comparison'' (STC).\cite{ishak2015simulation} These methods estimate a model for the conditional outcome expectation and average predictions over the target covariate distribution. G-computation exhibits increased precision relative to weighting, particularly when overlap is poor,\cite{remiro2022parametric} but relies on model-based extrapolation and can be prone to bias under model misspecification.\cite{vo2023cautionary} 

Weighting and G-computation are generally ``singly robust'': weighting, in most cases, depends on correct propensity score model specification; G-computation on correct outcome model specification. Decision-makers have expressed a preference for ``doubly robust'' estimation approaches that can consistently estimate the treatment effect as long as either the propensity score model or the outcome model is correct, but not necessarily both.\cite{phillippo2016nice,EUnetHTA2022practical, phillippo2018methods, vanier2024rapid} These methods should reduce the risk of bias by offering two opportunities for correct model specification. Despite this, doubly robust methods have rarely been applied to externally controlled SATs\cite{siu2024framework} and, to our knowledge, never to unanchored ITCs, despite recommendations from HTA agencies.\cite{phillippo2016nice, phillippo2018methods} 

One barrier may be a misunderstanding that MAIC is always doubly robust. MAIC enables consistent estimation when an implicit propensity score model is misspecified, but only if the true outcome model is linear with respect to balanced covariate functions -- termed ``linearly doubly robust''.\cite{josey2021transporting, cheng2023double, zhao2017entropy} In practice, outcomes rarely vary linearly with covariates.  Doubly robust methods for ITCs that are not necessarily restricted to linear outcome models are yet to be developed, with Josey et al (2021) recently identifying this as a research priority.\cite{josey2021transporting}

This paper clarifies existing approaches for doubly robust estimation in externally controlled SATs and proposes a doubly robust augmented MAIC estimator for unanchored ITCs. Section \ref{sec2} introduces the target estimand. Section \ref{sec3} outlines available estimators and our proposed approach. Section \ref{sec4} presents simulation study results comparing our proposed doubly robust augmented estimators against existing singly robust and other augmented estimators.  Lastly, we illustrate the application of the methods in an example analysis in Section \ref{sec5}, and conclude in Section \ref{sec6}.

\section{Estimands}\label{sec2}

We begin by defining the \textit{estimands} that can be targeted by externally controlled SATs. An estimand is a precise definition of the treatment effect, which should align with the clinical question of interest, the
research design and the analytical approach. The International Council of Harmonisation E9 (R1) Addendum, adopted by the FDA and EMA, specifies five estimand attributes: population, treatment(s), endpoint, summary effect measure, and strategies for intercurrent events.\cite{polito2024applying} We focus on the ``population'' and ``summary effect measure'' when defining the estimands, which are:

\begin{itemize}
\item The average treatment effect (ATE) among the combined SAT and external control;
\item The average treatment effect in the treated (ATT), among those participating in the SAT; and
\item The average treatment effect in the control (ATC), among the external control group. 
\end{itemize}
The difference between these summary effect measures is driven by them targeting different (sub) populations or applying to different ``analysis sets''. Having assumed the SAT and external control are random samples of their underlying target populations, we make no further distinction between sample-level and population-level estimands.

Using potential outcomes notation, let $Y^t$ represent the outcome under intervention $T=t$, with $t \in \{0, 1\}$, where $T=1$ denotes the SAT intervention (data source $S=1$) and $T=0$ the external control (data source $S=0$). Two potential outcomes, $(Y^1, Y^0)$, exist for every subject; one is observed, the other counterfactual. The ATE is:
\begin{equation*}
\textrm{ATE} =  g \left (\textrm{E} \left (Y^1 \right ) \right ) - g \left(\textrm{E} \left (Y^0 \right )  \right ),
\end{equation*}
where  the link function $g(\cdot)$ transforms potential outcome means into the plus/minus infinity range, and expectations are over the distribution of potential outcomes in the combined SAT and external control population. For binary outcomes, suitable links include identity, log or logit, to produce a risk difference, log
relative risk or log-odds ratio, respectively, as the summary effect measure. The ATT is:
\begin{equation*}
\textrm{ATT} =  g \left (\textrm{E} (Y^1 \mid S=1) \right ) - g \left(\textrm{E} \left (Y^0 \mid S=1 \right )  \right ),
\end{equation*}
with expectations taken over the SAT (sub) population. The ATC is:
\begin{equation*}
\textrm{ATC} =  g \left (\textrm{E} (Y^1 \mid S=0) \right ) - g \left(\textrm{E}(Y^0 \mid S=0)  \right ),
\end{equation*}
with expectations over the external control (sub) population.

Within RCTs, the ATE, ATT and ATC are identical in expectation. However, they generally differ in externally controlled SATs, and will almost invariably do so where there is treatment effect heterogeneity by the covariates, i.e., effect measure modification. We view the ATE target population, defined by pooling the SAT and the external control, as somewhat
ambiguous in this context. As such, the target estimand in an externally controlled SAT is often either the ATT or the ATC. 
 
The ATT is typically the primary estimand for regulatory drug approval, consistent with emulating a randomized comparison in the pivotal trial population, with the external control mimicking the internal comparator arm of a registrational clinical trial. The ATT is also compatible with the mean absolute outcome that is targeted by the SAT, $\textrm{E}(Y^1 \mid S=1)$, preserving the original SAT results. Nevertheless, SAT populations are often highly selected and may lack ``real-world'' representativeness, making the ATT potentially less appealing for HTA, where generalizability to routine clinical practice is a priority. 

The ATC can be more desirable for external validity, as natural history and RWD-based external controls have relatively broad inclusion criteria and heterogeneous target populations. However, external controls based on historical trials will not reflect the current standard of care and RWD-derived controls are often country-specific, not necessarily transferable to the relevant jurisdiction for decision-making.  Sample size considerations also influence the estimand choice. Both SATs and external controls often have low sample sizes and covariate adjustment may reduce effective sample sizes further. Consider weighting, where the estimand impacts the definition of the weights. Targeting the ATT implies preserving the original SAT, re-weighting and reducing the effective sample size of the external control. Conversely, targeting the ATC implies the reverse.

Finally, where IPD are available for the SAT but only AD for the external control, as in unanchored ITCs, the ATC is often targeted by necessity.\cite{phillippo2016nice, phillippo2018methods, lambert2023enriching} Throughout this manuscript, we assume unlimited subject-level data access but target the ATC. Our methodological approaches are also applicable where subject-level data are unavailable for the external control, and to target the ATT instead of the ATC, with caveats discussed in Section \ref{subsec38} and the Supplementary Material, respectively.

\section{Methodology}\label{sec3}

\subsection{Data and assumptions}\label{subsec31}

As per Section \ref{sec2}, let $T=t$ denote a time-fixed binary treatment, with $t\in \{0, 1\}$, such that $T=1$ represents the active intervention and $T=0$ the control. Let $S=s$ denote the data source, with $s \in \{0, 1\}$, such that $S=1$ represents the SAT and $S=0$ the external data source. In addition, let $\mathbf{X}$ denote vector-valued pre-treatment baseline covariates, e.g., clinical or demographic characteristics, measured across the SAT and the external data source. Let $Y$ denote the clinical outcome of interest. We assume that only distributional differences in $\mathbf{X}$ are preventing exchangeability between the SAT and external subjects, and that covariates and outcomes are defined and measured similarly across data sources. 

The observed IPD consist of $(S_i, \mathbf{X}_i, T_i, Y_i)$,  $i=1, \dots, n_1+n_0$, realizations of $(S, \mathbf{X}, T, Y)$ denoting the data source, baseline covariates, treatment assignment and observed outcome for subject $i$. Here, the SAT and external data source have been stacked, with $n_1$ and $n_0$ as the sample sizes of the SAT and the external data source, respectively. It is assumed that all individuals in the SAT are under $T=1$ and all individuals in the external data source are under $T=0$, such that the control group is fully external.  To be clear, we have $S_i=T_i$ for all $i=1, \dots, n$, where $n = n_1+n_0$; and also have that $S_i=1$ and $T_i=1$ for all $i=1,\dots,n_1$, and  $S_i=0$ and $T_i=0$ for all $i=n_1+1, \dots, n$. We shall assume that there is no missingness or measurement error. 

The observed outcome for subject $i$ is $Y_i = Y_i^1T_i + Y_i^0(1-T_i)$, where $Y_i^t$ is the potential outcome had subject $i$ been assigned treatment $t \in \{0, 1\}$, with $Y_i=Y_i^1$ if $i=1,\dots, n_1$ and $Y_i=Y_i^0$ if $i=n_1+1, \dots, n$. Namely, the observed outcome for an individual in the SAT equals their potential outcome under the active intervention, and the observed outcome for an individual in the external data source equals their potential outcome under the control. Implicit in the notation is the stable unit treatment value assumption (SUTVA): that there is no interference between subjects and there is treatment version irrelevance, i.e., one well-defined version of the active intervention and the control across all subjects and data sources.\cite{zhou2024causal} Also implicit is that there is no direct effect of trial participation.\cite{zhou2024causal} Namely, that trial participation -- in the SAT or a historical trial, for that matter -- does not affect the outcome except through treatment assignment itself, i.e., there are no Hawthorne effects.\cite{braunholtz2001randomized, dahabreh2019hawthorne}  

To estimate the ATC, we must construct estimators for $\mu^1_0 = \textrm{E}(Y^1 \mid S=0)$ and $\mu^0_0= \textrm{E}(Y^0 \mid S=0)$. Outcomes for the subjects from the external data source have been generated under the control and we  assume that there is no informative missingness or measurement error. Hence, unbiased estimation of $\mu^0_0$ should be trivial using the sample mean, such that $\hat{\mu}^0_0= \frac{1}{n_0}\sum_{i=n_1+1}^nY_i$. Conversely, while the active intervention has been investigated in the SAT, its outcomes in the external control (sub) population are unobserved. Our challenge is therefore to produce a reliable estimate $\hat{\mu}^1_0$ of the mean absolute outcome $\mu^1_0$ under the active intervention in the external control (sub) population, based on the observed data. 

Two causal identification conditions, together known as \textit{strong ignorability}, are required to construct a valid estimator of $\mu_0^1$. These ensure that the SAT and external control outcomes are comparable given adjustment for baseline covariates. The first assumption is \textit{conditional data source ignorability}; formally, $Y_i^1 \ind S_i \mid \mathbf{X}_i$ for all $i=1,\dots, n$. Namely, conditional on baseline covariates, the potential outcome under the active intervention is independent of the data source. This is akin to the conditional constancy or exchangeability of absolute outcomes invoked for unanchored ITCs, used to transport mean absolute outcomes under $T=1$ from $S=1$ to $S=0$.\cite{phillippo2016nice, phillippo2018methods} Conditional ignorability is a strong assumption, resting on the SAT and the external control capturing all variables that are prognostic of outcome under the active intervention. 


The second assumption is \textit{positivity} or \textit{overlap}. That is, the support of the baseline covariates in the external control is contained within that of the SAT. Mathematically, the probability of SAT participation, conditional on the covariates necessary to ensure ignorability, should be bounded away from zero and one: $0 < \textrm{Pr}(S=1 \mid \mathbf{X}=\mathbf{x}) < 1$ for all $\mathbf{x}$ with positive density in the external control, i.e., for all $\mathbf{x}$ such that $f(\mathbf{x} \mid S=0)>0$. Hence, it is possible to have SAT subjects in all regions of the covariate space in $S=0$.\cite{zhou2024causal, dahabreh2019generalizing, buchanan2018generalizing} Positivity violations can be deterministic or random. The former arise structurally, due to non-overlapping SAT and external control eligibility criteria. The latter arise empirically due to chance, particularly with small sample sizes.\cite{westreich2010invited} To enforce positivity, analysts may subset the SAT based on the selection criteria of the external control.\cite{phillippo2016nice, phillippo2018methods} However, this further reduces the sample size of the SAT. Positivity is typically assessed by comparing the empirical distributions of the covariates in the SAT and the external control.\cite{glimm2022geometric} While outcome modeling-based approaches such as G-computation can overcome failures of positivity, they do so by potentially problematic and difficult-to-diagnose model-based extrapolation. Even minor model misspecification over the observed covariate space in the SAT may lead to poor extrapolation in unobserved regions of the covariate space.\cite{vo2023cautionary}  

Analogously, targeting the ATT would require constructing estimators for $\mu^1_1 = \textrm{E}(Y^1 \mid S=1)$ and $\mu^0_1= \textrm{E}(Y^0 \mid S=1)$. Here, the challenge is the estimation of $\mu^0_1$ because outcomes under the control have not been generated in the SAT. The conditional ignorability assumption would formally be $Y_i^0 \ind S_i \mid \mathbf{X}_i$ for all $i=1,\dots, n$, and would rest on the SAT and the external control measuring all variables that are prognostic of outcome under the control. The positivity assumption would be $0 < \textrm{Pr}(S=0 \mid \mathbf{X}=\mathbf{x}) < 1$ for all $\mathbf{x}$ with positive density in the SAT, $f(\mathbf{x} \mid S=1)>0$, such that the support of the baseline covariates in the SAT is contained within that of the external control and it is possible to have external control subjects in all regions of the SAT covariate distribution. In this setting, analysts may apply the SAT selection criteria to the external control to guarantee that there is sufficient overlap. While the methods in the next sections target the ATC, we present any required modifications to target the ATT in the Supplementary Material. 

\subsection{Inverse odds weighting}\label{subsec32}

We first present a covariate adjustment method that models the data source assignment mechanism, conditional on baseline covariates, to estimate weights.\cite{ren2023comparing, loiseau2022external, josey2021transporting, dahabreh2020extending, colnet2024causal} Where the target estimand is the ATC, SAT subjects are weighted by their inverse conditional odds of SAT participation -- their conditional odds of external control participation -- to transport the SAT outcomes to the external control (sub) population. Such ``inverse odds'' weights (IOW) are defined as: 
\begin{equation}\label{IO_weights}
w_i 
=
\frac{(1-e_i)S_i}
{e_i} + (1-S_i),
\end{equation}
for subject $i=1,\dots, n$, where the propensity score $e_i=e(\mathbf{X}_i)=\textrm{Pr}(S_i=1 \mid \mathbf{X}_i)$ denotes the conditional probability of SAT participation given covariates $\mathbf{X}_i$ for subject $i$. In Equation \ref{IO_weights}, note that the SAT subjects $(S_i=1)$ are weighted as $w_i=(1-e_i)/e_i$, whereas the external control subjects $(S_i=0)$ are unweighted, i.e., assigned a weight of $w_i=1$. 

In practice, the true propensity scores are unknown. Almost invariably, there are multiple baseline covariates and at least one of these is continuous, such that a data source assignment model is required to estimate the propensity scores. The model is often a logistic regression:
\begin{equation}
\logit(e_i) = \alpha_0 + \mathbf{c}(\mathbf{X}_i)^\top \mathbf{\alpha},
\label{logistic_PS}
\end{equation}
where $\logit(e_i)=\ln \left ((e_i)/(1-e_i) \right)$, $\alpha_0=\ln \left (\textrm{Pr}(S_i=1|\mathbf{c}(\mathbf{X}_i)=0)/(\textrm{Pr}(S_i=0|\mathbf{c}(\mathbf{X}_i)=0) \right)$ is an intercept term, $\mathbf{\alpha}$ is a vector of regression parameters, and $\mathbf{c}(\mathbf{X}_i)= [c_1(\mathbf{X}_{i}), c_2(\mathbf{X}_{i}), \dots, c_k(\mathbf{X}_i)]^\top$ is a vector of covariate ``balance functions'' for subject $i=1,\dots n$. This is the set of functions containing the distributional features to be balanced between the SAT and the external control,\cite{josey2021transporting, josey2022calibration} potentially including sensible transformations of the covariates, e.g., polynomials and interaction terms.

The logistic regression is typically fitted to the concatenated IPD using maximum-likelihood estimation, with the regression coefficient point estimates denoted by $\hat{\alpha}_0$ and $\hat{\mathbf{\alpha}}$, and model-based propensity scores for subject $i=1, \dots, n_1$, predicted by $\hat{e}_i=\logit^{-1}\left( \hat{\alpha}_0 +
\mathbf{c}(\mathbf{X}_i)^\top \hat{\mathbf{\alpha}} \right)=\expit \left ( \hat{\alpha}_0 +
\mathbf{c}(\mathbf{X}_i)^\top \hat{\mathbf{\alpha}} \right)$, where $\expit(\cdot)=\exp(\cdot)/\left(1 + \exp(\cdot) \right)$. Weight estimates $\hat{w}_i$ for $i=1,\dots, n_1$ are derived by plugging the corresponding propensity score predictions into Equation \ref{IO_weights}. With correct specification of the model in Equation \ref{logistic_PS}, such that the log-odds of SAT participation are linear across the balance functions of the covariates, $\hat{e}_i$ and $\hat{w}_i$ consistently estimate the true conditional probability and inverse odds of SAT participation, respectively.

The ATC is estimated by contrasting the weighted average of observed outcomes under the active intervention with the unweighted average of observed outcomes for the external control. As per Section \ref{sec2}, mean absolute outcomes are converted to the additive scale imposed by link function $g(\cdot)$ prior to taking the difference between treatments on such scale, leading to the \textbf{IOW} estimator for the ATC:
\begin{equation}
\widehat{\textrm{ATC}} 
= 
g 
\underbrace
{
\left ( 
\frac{1}{n_0}\sum_{i=1}^{n_1} \hat{w}_i Y_i
\right )
}_{\hat{\mu}_0^1}
-
g 
\underbrace
{
\left (
\frac{1}{n_0}\sum_{i=n_1+1}^n Y_i
\right )
}_{\hat{\mu}_0^0},
\label{norm_weights_1}
\end{equation}
The mean absolute outcome estimate for the active intervention can be bounded within its feasible range, e.g., between 0 and 1 for probabilities, by normalizing or ``stabilizing'' the weights so that they sum to one.\cite{dahabreh2019generalizing, dahabreh2020extending} This results in the alternative \textbf{normalized IOW} estimator for the ATC:
\begin{equation}
\widehat{\textrm{ATC}} 
= 
g 
\underbrace
{
\left ( 
\frac{ \sum_{i=1}^{n_1} \hat{w}_i Y_i}{\sum_{i=1}^{n_1}  \hat{w}_i}
\right )
}_{\hat{\mu}_0^1}
-
g 
\underbrace
{
\left (
\frac{1}{n_0}\sum_{i=n_1+1}^n Y_i
\right )
}_{\hat{\mu}_0^0},
\label{norm_weights_2}
\end{equation}
which should provide improved finite sample properties and more stable and precise estimation.\cite{busso2014new, hernan2000marginal} Drawing an analogy with survey sampling, the estimator in Equation \ref{norm_weights_1} is a Horvitz-Thompson-type estimator and that in Equation \ref{norm_weights_2} is a Hajek-type estimator, with the latter typically considered to improve the performance of the former in the literature.\cite{sarndal2003model} 

In expectation, if the model in Equation \ref{logistic_PS} is correctly specified, the estimated weights, ($\hat{w}_{i}$, for $i$ in 1,...,$n_{1}$) will balance the covariate distribution of the SAT with respect to that of the external control, enabling consistent estimation of mean absolute outcome $\mu^1_0$ and the ATC.  A mathematical derivation showing the consistency of the IOW estimators is provided in the Appendix.

Unfortunately, this ``modeling approach'' to weighting,\cite{chattopadhyay2020balancing, filla2024balancing} where propensity scores are explicitly modeled as a function of baseline covariates by a logistic regression, then estimated by maximizing the fit of such regression, has certain limitations:\cite{chattopadhyay2020balancing, filla2024balancing}

\begin{itemize}
\item The resulting weights do not produce adequate covariate balance if the propensity score model is misspecified, and even a correctly specified model does not guarantee balance in finite samples;

\item Propensity score predictions that are close to zero produce extreme and highly variable weights, which lead to  unstable and imprecise ATC estimation, particularly where overlap is poor or the sample size of the SAT is small; and

\item There is limited applicability when covariate IPD for the external control are unavailable and only marginal summary moments from published tables of baseline characteristics are available. 
\end{itemize}

\subsection{Entropy balancing (matching-adjusted indirect comparison)}\label{subsec33}

The limitations of the weighting methods in Section \ref{subsec32} motivate alternative ``balancing'' or ``calibration'' approaches to weighting.  These estimate weights under the condition that covariates are balanced, viewing balance as an optimization problem, without explicitly modeling the propensity score. Generally, balancing approaches to weighting are: (1) less susceptible to bias by directly enforcing covariate balance; (2) produce more stable weights, which translate into larger effective sample sizes and more precise treatment effect estimation; and (3) are applicable where only aggregate-level marginal covariate moments are available for the external control.\cite{chattopadhyay2020balancing, filla2024balancing, hirshberg2017two, wang2021matching} 

Our focus here is on an entropy balancing approach\cite{hainmueller2012entropy} called matching-adjusted indirect comparison (MAIC),\cite{signorovitch2010comparative, cheng2020statistical, josey2021transporting} but see Chattopadhyay et al (2020) and Filla et al (2024) for details about similar balancing techniques.\cite{chattopadhyay2020balancing, filla2024balancing} MAIC is the most popular balancing method in the context of externally controlled SATs. It has many features that are considered desirable: ``linear double robustness'', minimally dispersed weights, and the estimation of odds weights that are guaranteed to be positive, resulting in increased interpretability and sample-boundedness, i.e., interpolating the observed data as opposed to extrapolating beyond its support.\cite{filla2024balancing, tan2010bounded} We review the main steps of MAIC, building on prior literature.\cite{signorovitch2010comparative,wang2021matching,phillippo2020equivalence, remiro2022two, jackson2021alternative, jiang2024comprehensive} 

While MAIC does not explicitly model the propensity score as a function of baseline covariates, it implicitly assumes the following logistic model for data source assignment:
\begin{equation}
\ln(v_i) 
\propto \ln  \Bigg ( \frac{(1-e_i)}{e_i} \Bigg )
= \gamma_0 + 
\mathbf{c}(\mathbf{X}_i)^\top \mathbf{\gamma},
\label{maic_model} 
\end{equation}
where $v_i$ is a weight proportional to the inverse conditional odds of SAT participation for subject $i=1,\dots, n$, $\gamma_0$ is an intercept term parameter and $\mathbf{\gamma}$ is a vector of model parameters. In Equation \ref{maic_model}, it is the log-odds of external control participation, $\logit(1-e_i)$, that are linear across the covariate balance functions. Because $\logit(1-e_i)=-\logit(e_i)$, this implies that the log-odds of SAT participation, $\logit(e_i)$, also vary linearly with $\mathbf{c}(\mathbf{X}_i)$ as per Equation \ref{logistic_PS}. 


Signorovitch et al\cite{signorovitch2010comparative} propose using the ``method of moments'' to estimate the model in Equation \ref{maic_model}, such that:  
\begin{align}
\frac{
\sum_{i=1}^{n_1} v_i \mathbf{c}(\mathbf{X}_i)  
}
{
\sum_{i=1}^{n_1} v_i 
}
&=
\frac{1}{n_0}
\sum_{i=n_1+1}^{n}
\mathbf{c}(\mathbf{X}_i) 
\label{balancing_constraint}
\\
\frac{
\cancel{\exp \left  (\gamma_0 \right )}
\sum_{i=1}^{n_1} 
\exp \left (\mathbf{c}(\mathbf{X}_i)^\top \mathbf{\gamma} \right ) \mathbf{c}(\mathbf{X}_i)  
}
{
\cancel{\exp \left  (\gamma_0 \right )}
\sum_{i=1}^{n_1} 
\exp \left (\mathbf{c}(\mathbf{X}_i)^\top \mathbf{\gamma} \right )
}
&=
\underbrace
{
\frac{1}{n_0}
\sum_{i=n_1+1}^{n}
\mathbf{c}(\mathbf{X}_i)
}_{\hat{\mathbf{\theta}}}
\label{balancing_constraint_2}
\\
\frac{
\sum_{i=1}^{n_1} 
\exp \left (\mathbf{c}(\mathbf{X}_i)^\top \mathbf{\gamma} \right ) \mathbf{c}(\mathbf{X}_i)  
}
{
\sum_{i=1}^{n_1} 
\exp \left (\mathbf{c}(\mathbf{X}_i)^\top \mathbf{\gamma} \right )
}
&=
\hat{\mathbf{\theta}}.
\label{balancing_constraint_3}
\end{align}
where $\hat{\mathbf{\theta}} = (
\hat{\mathbf{\theta}}_1, \hat{\mathbf{\theta}}_2,\dots,\hat{\mathbf{\theta}}_k)^\top$ is a vector of covariate balance function moments $j=1,\dots, k$ for the external control sample, with $\hat{\mathbf{\theta}}_j = \frac{1}{n_0}\sum_{i=n_1+1}^n c_j(\mathbf{X}_i)$ assumed to be a consistent estimator for
$\mathbf{\theta}_j$. Equation \ref{balancing_constraint} is a constraint enforcing that the covariate distributional features of the weighted SAT subjects are exactly balanced with respect to those of the unweighted external control subjects. Equation \ref{balancing_constraint_2} follows from introducing the assumed model in Equation \ref{maic_model} into the balancing constraint, and Equation \ref{balancing_constraint_3} results from the exponentiated intercept terms canceling out.   

Replacing $\mathbf{\gamma}$ with estimate $\hat{\mathbf{\gamma}}$ in Equation \ref{balancing_constraint_3} and centering the SAT covariate balance functions on their external control means, one obtains: 
\begin{equation}
\frac{
\sum_{i=1}^{n_1} 
\exp \left (\mathbf{c^{*}}(\mathbf{X}_i)^\top \hat{\mathbf{\gamma}} \right ) \mathbf{c^{*}}(\mathbf{X}_i)  
}
{
\sum_{i=1}^{n_1} 
\exp \left (\mathbf{c^{*}}(\mathbf{X}_i)^\top \hat{\mathbf{\gamma}} \right )
}
=
\mathbf{0},
\label{centered_constraint}
\end{equation}
where $\mathbf{0}$ is a vector of zeros and $\mathbf{c^{*}}(\mathbf{X}_i)= \mathbf{c(\mathbf{X}_i)} - \hat{\mathbf{\theta}}$ is a vector of centered covariate balance functions for subject $i=1,\dots,n_1$ in the SAT. Then, because the denominator is positive, Equation \ref{centered_constraint} is equal to $\sum_{i=1}^{n_1} 
\exp \left (\mathbf{c^{*}}(\mathbf{X}_i)^\top \hat{\mathbf{\gamma}} \right ) \mathbf{c^{*}}(\mathbf{X}_i)  
=
\mathbf{0}$. Solving for $\hat{\mathbf{\gamma}}$ is equivalent to minimizing the objective function:
\begin{equation}
Q(\hat{\mathbf{\gamma}}) =
\sum_{i=1}^{n_1} \exp \left ( \mathbf{c^{*}}(\mathbf{X}_i)^\top \hat{\mathbf{\gamma}} \right ),
\label{objective_function}
\end{equation}
as the derivative of $Q(\hat{\mathbf{\gamma}})$ with respect to $\hat{\mathbf{\gamma}}$ is $\sum_{i=1}^{n_1} 
\exp \left (\mathbf{c^{*}}(\mathbf{X}_i)^\top \hat{\mathbf{\gamma}} \right ) \mathbf{c^{*}}(\mathbf{X}_i)$. The objective function in Equation $\ref{objective_function}$ is strictly convex and can be minimized using standard Newton-type convex optimization algorithms,\cite{hainmueller2012entropy} yielding an unique finite solution corresponding to the global minimum of $Q(\hat{\mathbf{\gamma}})$. We have $v_i \propto \exp \left ( \mathbf{c} (\mathbf{X}_i)^\top \mathbf{\gamma} \right ) \propto 
\exp 
\left ( 
\left (
\mathbf{c}(\mathbf{X}_i)-
\mathbf{\theta}
\right )
^\top \mathbf{\gamma} \right )
=
\exp
\left ( \mathbf{c^{*}}(\mathbf{X}_i)^\top \mathbf{\gamma} \right )
$. Subject to the normalization constraint $\sum_{i=1}^{n_1} \hat{v}_i=1$, weights for each individual $i=1,\dots,n_1$ in the SAT are estimated as:

\begin{equation}
\hat{v}_i = 
\frac{
\exp \left ( \mathbf{c^{*}}(\mathbf{X}_i)^\top \hat{\mathbf{\gamma}} \right )
}
{
\sum_{i=1}^{n_1}
\exp \left ( \mathbf{c^{*}}(\mathbf{X}_i)^\top \hat{\mathbf{\gamma}} \right )
}.
\label{weight_equation}
\end{equation}
Note that this definition of the entropy balancing weights coincides with the definitions presented by Jiang et al (2024) \cite{jiang2024comprehensive} and by Jackson et al (2021).\cite{jackson2021alternative}

Similar to Equation \ref{norm_weights_2}, the \textbf{MAIC} estimator for the ATC contrasts absolute outcomes on the additive scale:
\begin{equation}
\widehat{\textrm{ATC}} 
= 
g 
\underbrace
{
\left ( \sum_{i=1}^{n_1} \hat{v}_i Y_i
\right )
}
_{\hat{\mu}_0^1}
-
g
\underbrace
{
\left (
\frac{1}{n_0}\sum_{i=n_1+1}^n Y_i
\right )
}
_{\hat{\mu}_0^0}
,
\label{norm_weights_3}
\end{equation}
where the weights have already been normalized to sum to one. Alternatively, fitting a weighted univariable regression of outcome on treatment to the concatenated IPD (with weights $\hat{v}_i$ for $i=1,\dots,n_1$ and unit weights for $i=n_1+1,\dots, n$) has been proposed, with the treatment coefficient of the fitted model yielding an ATC estimate.\cite{ishak2015simulation,
park2024unanchored,jiang2020performance, hatswell2020effects, chandler2024uncertain} We discourage this approach. The weights estimated by Equation \ref{weight_equation} are relative; their arbitrary rescaling by a constant of proportionality, e.g., a normalization constant, will also balance the specified covariate functions and does not affect $\hat{\mu}_0^1$ and $\widehat{\textrm{ATC}}$ in Equation \ref{norm_weights_3}.\cite{phillippo2020equivalence,jackson2021alternative} Nevertheless, rescaling the weights while retaining weights of one for the external control subjects results in different fitted models.\cite{jackson2021alternative}

The method of moments MAIC estimator gives the Lagrangian dual solution to an entropy balancing primal problem: minimizing the negative entropy of the weights.\cite{josey2021transporting, cheng2023double, phillippo2020equivalence} Namely, minimizing the objective function in Equation \ref{objective_function} is equivalent to minimizing the negative entropy $\sum_{i=1}^{n_1} v_i \ln (v_i)$ with Lagrange multipliers.\cite{cheng2023double, phillippo2020equivalence} The dual optimization problem is easier to solve than the primal\cite{hainmueller2012entropy} and has been formulated in different ways.\cite{josey2021transporting,phillippo2020equivalence, jiang2024comprehensive} These may perform differently computationally -- e.g., if minimization is performed on the log scale\cite{phillippo2020equivalence} -- but result in the same unique dual solution, up to numerical error, due to strict convexity. Because the negative entropy measures the distance of the weights from a uniform distribution, its minimization should produce less disperse weights than the modeling approach in Section \ref{subsec32}.  

MAIC (entropy balancing) has a  ``linear double robustness'' property as it is consistent under two distinct underlying data-generating models:\cite{josey2021transporting,cheng2023double,zhao2017entropy} (1) the true log-odds of the propensity score are linear across the specified balance functions; \textit{or} (2) the true potential outcome under the active intervention is linear across the specified balance functions. Respectively, either $\logit(e_i) = \alpha_0 + \mathbf{c}(\mathbf{X}_i)^\top \mathbf{\alpha}$ for some parameters $\alpha_0$ and $\mathbf{\alpha}$; \textit{or}, $\textrm{E}(Y^1_i \mid \mathbf{X}_i) = \beta_0 + \mathbf{c}(\mathbf{X}_i)^\top \mathbf{\beta}$ for some parameters $\beta_0$ and $\mathbf{\beta}$.

For continuous covariates, it is common practice to only balance first-order sample moments (means) by setting $\mathbf{c}(\mathbf{X}_i)=c_1(\mathbf{X}_i)=\mathbf{X}_i$ and $\hat{\mathbf{\theta}} = 
\hat{\mathbf{\theta}}_1 = \frac{1}{n_0}\sum_{i=n_1+1}^n \mathbf{X}_i$, in which case consistency is guaranteed as long as either the true propensity score model is $\logit(e_i) = \alpha_0 + \mathbf{X}_i^\top \mathbf{\alpha}$; \textit{or}, the potential outcome for the active intervention varies linearly with the covariates $\mathbf{X}_i$ that are mean-balanced: $\textrm{E}(Y^1_i \mid \mathbf{X}_i) = \beta_0 + \mathbf{X}_i^\top \mathbf{\beta}$. An alternative strategy is to also balance second-order sample moments (variances) by setting $\mathbf{c}(\mathbf{X}_i)=[c_1(\mathbf{X}_{i}), c_2(\mathbf{X}_{i})]^\top=[\mathbf{X}_{i},\mathbf{X}^2_{i}]^\top$, $\hat{\mathbf{\theta}} = 
(\hat{\mathbf{\theta}}_1, \hat{\mathbf{\theta}}_2)^\top$, $\hat{\mathbf{\theta}}_1 = \frac{1}{n_0}\sum_{i=n_1+1}^n \mathbf{X}_i$ and $\hat{\mathbf{\theta}}_2 = \frac{1}{n_0}\sum_{i=n_1+1}^n \mathbf{X}_i^2$, in which case consistency is achieved if either the true propensity score model is $\textrm{logit}(e_i) = \alpha_0 + \mathbf{X}_i^\top \mathbf{\alpha_{1}} + (\mathbf{X}_i^{2})^\top \mathbf{\alpha_{2}}$; \textit{or}, the potential outcome for the active intervention varies quadratically with $\mathbf{X}_i$, such that $\textrm{E}(Y^1_i \mid \mathbf{X}_i) = \beta_0 + \mathbf{X}_i^\top \mathbf{\beta}_1 + (\mathbf{X}_i^2)^\top \mathbf{\beta}_2$. Balancing the means of the covariates and the squared covariates enforces that variances are balanced because $\textrm{Var}(X) = \textrm{E}(X^2) - \textrm{E}(X)^2$.

One could conjecture more flexible implicit data-generating models by considering other non-linear transformations of $\mathbf{X}_i$, e.g., higher-order polynomial terms and flexible basis functions such as splines, and balancing on the means of the transformed covariates. Moreover, one could go beyond the marginal moments of individual covariates and also balance joint covariate distributions; for instance, by including quantiles of interactions for pairs or triples of covariates.\cite{zubizarreta2015stable} However, pursuing these more ambitious balancing strategies is often infeasible:

\begin{itemize}
\item As the number of balancing conditions increases, it is more likely that $\mathbf{\theta}$ falls outside the convex hull of $\mathbf{c}(\mathbf{X}_i)$ for $i=1,\dots,n_1$.\cite{glimm2022geometric} This implies that a feasible weighting solution to the convex optimization problem does not exist: there is no set of positive weights that can enforce balance in the required distributional features and the numerical optimization algorithm will fail to converge.\cite{chattopadhyay2020balancing} 
\item Increasing the number of balancing conditions leads to further reductions in effective sample size and precision, which are particularly problematic with low sample sizes and poor covariate overlap. 
\item Where covariate IPD for the external control are unavailable, only first- and second-order marginal moments from published tables of baseline characteristics are often available for balancing. Higher-order moments and the means of transformed covariates are rarely reported. 
\end{itemize}

\subsection{G-computation}\label{subsec34}

An alternative to weighting involves postulating a model $m \left(\mathbf{X}_i; \mathbf{\beta} \right )$ for the potential outcome expectation under the active intervention, conditional on covariates $\mathbf{X}_i$:
\begin{align}
q\left(
\textrm{E}(Y_i^1 \mid \mathbf{X}_i; \mathbf{\beta})
\right)=m \left(\mathbf{X}_i; \mathbf{\beta} \right ),
\label{outcome_model}
\end{align}
where $q(\cdot)$ denotes an appropriate link function and $\mathbf{\beta}$ is a vector of model parameters encoding the covariate-outcome relationships. For instance, a logistic regression could be used for binary outcomes, such that the model is binomial, the link function is the logit and the potential outcome expectation is on the probability scale. We have assumed that the model for the conditional outcome expectation is parametric, but this need not necessarily be the case.\cite{kern2016assessing} 

The model in Equation \ref{outcome_model} is fitted to the SAT participants $i=1,\dots, n_1$ using maximum-likelihood estimation, such that the fitted model $m \left (\mathbf{X}_i; \hat{\mathbf{\beta}} \right)$ with parameter estimates $\hat{\mathbf{\beta}}$ is an estimator of the (transformed) conditional expectation $q\left(\textrm{E}(Y_i^1 \mid \mathbf{X}_i, S_i=1)\right)$. Based on the model, potential outcomes under the active intervention are predicted for each subject in the external control: 
\begin{equation}
\hat{Y}_i^1 = q^{-1} \left (m({\mathbf{X}_i};\hat{\mathbf{\beta}})\right).    
\label{outcome_prediction}
\end{equation}
In this case, $i=n_1+1,\dots,n$, and the predicted outcomes are counterfactual because the subjects in the external control have not received the active intervention. 

By averaging (``marginalizing'') the potential outcome predictions generated by Equation \ref{outcome_prediction} over the empirical covariate distribution of the external control, one obtains the \textbf{G-computation} estimator,\cite{loiseau2022external, wang2017g, robins1986new, keil2014parametric, remiro2022parametric} also known as ``regression standardization'' or ``plug-in G-formula'', for the mean potential outcome $\mu_0^1$ had subjects in the external control received the active intervention:
\begin{equation}
\hat{\mu}_0^1= 
\frac{1}
{n_0}
\sum_{i=n_1+1}^{n}
\hat{Y}_i^1.
\label{Gcomp_absolute}
\end{equation}
Consequently, mean absolute outcomes are converted to the additive scale imposed by link function $g(\cdot)$, prior to taking the difference between treatments on such scale to derive the G-computation estimator for the ATC:
\begin{equation}
\widehat{\textrm{ATC}} 
= 
g 
\underbrace
{
\left (
\frac{1}
{n_0}
\sum_{i=n_1+1}^{n}
\hat{Y}_i^1
\right )
}_{\hat{\mu}_0^1}
-
g 
\underbrace
{
\left (
\frac{1}{n_0}\sum_{i=n_1+1}^n Y_i
\right )
}_{\hat{\mu}_0^0}.
\label{GComp_ATC}
\end{equation}
Note that the link function $g(\cdot)$ used for summarizing the treatment effect does not necessarily need to match the link function $q(\cdot)$ used for modeling.\cite{remiro2024transportability} 

The readership may be more familiar with the well-known G-computation estimator for the ATE, which contrasts averages of \textit{predicted} potential outcomes between both treatment groups.\cite{ren2023comparing} Our G-computation estimator in Equation \ref{GComp_ATC} contrasts the average of predicted counterfactual outcomes under the active intervention with the average of \textit{observed} factual outcomes for the external control group. To estimate the ATC, only potential outcomes for the SAT subjects need to be predicted because all potential outcomes under the control have already been observed for the subjects in the external control.\cite{wang2017g} 

The G-computation estimators in Equations \ref{Gcomp_absolute} and \ref{GComp_ATC} rely on the outcome model in Equation \ref{outcome_model} being correctly specified, in order to be consistent for the mean absolute outcome ${\mu}_0^1$ and the ATC, respectively. Assuming correct model specification for all estimators, G-computation is more precise and efficient than weighting, particularly when poor overlap leads to large reductions in effective sample size.\cite{loiseau2022external, remiro2022parametric} However, the increase in precision is achieved by implicit extrapolation into non-overlapping regions of the covariate space, hiding underlying failures of the positivity assumption.\cite{vo2023cautionary} Model misspecification bias is almost impossible to diagnose in extrapolated regions, and there is typically no inflation of the variance to reflect the extrapolation uncertainty. 

\subsection{Doubly robust augmented weighting estimators}\label{subsec35}

The limitations of the methods in Section \ref{subsec32}, Section \ref{subsec33} and Section \ref{subsec34} motivate the explicit augmentation of the weighting estimators,\cite{funk2011doubly} allowing for a less restrictive outcome model that permits non-linear link functions and/or outcomes that depend on non-linear covariate transformations. Our proposed augmented weighting estimators will not use the outcome model to extrapolate, but to gain bias-robustness -- and, potentially, some precision\cite{lunceford2004stratification} -- with respect to their weighting counterparts. To construct such augmented estimators, we proceed as follows. 




Firstly, based on Funk et al (2011) and Shinozaki and Matsuyama (2015),\cite{funk2011doubly, shinozaki2015brief} we propose combining the modeling approach to inverse odds weighting, described in Section \ref{subsec32}, with G-computation. Specifically, suppose we have fitted the outcome model in Equation \ref{outcome_model} to the SAT. We now let $\hat{Y_i}^1= q^{-1} \left ( m (\mathbf{X}_i; \hat{\mathbf{\beta}} ) \right )$ be a prediction of potential outcome $Y_i^1$ for the active intervention based on the fitted regression, not only for the external control subjects but for all subjects $i=1,\dots,n$ in the SAT and the external control. The propensity score weights $\hat{w}_i$ derived for $i=1,\dots, n_1$ are used to add an error-correcting term to the G-computation estimator. The resulting \textbf{doubly robust (DR) augmented IOW} estimator for the mean absolute outcome $\mu_0^1$ is:
\begin{align}
\hat{\mu}_0^1
=&
\frac{1}
{n_0}
\left (
\sum_{i=1}^{n_1}
\hat{w}
_i
\left (
Y_i
-
\hat{Y}^1_i
\right )
\right )
+
\frac{1}{n_0}
\sum_{i=n_1+1}^{n}
\hat{Y_i}^1 \nonumber \\
=&
\frac{1}
{n_0}
\left (
\sum_{i=1}^{n_1}
\hat{w}
_i
\epsilon_i^1
+
\sum_{i=n_1+1}^{n}
\hat{Y_i}^1
\right ),
\label{absolute_dr2}
\end{align}
where $\epsilon_i^1=Y_i -\hat{Y}_i^1$ is a residual term for subject $i=1,\dots,n_1$ in the SAT.  Note that this estimator exactly corresponds to the doubly robust estimator proposed by Shinozaki and Matsuyama (2015),\cite{shinozaki2015brief} except that their estimator is for $\textrm{E}(Y^{0}| S = 1)$ while the estimator in Equation \ref{absolute_dr2} is for $\textrm{E}(Y^{1}| S = 0)$. The estimand ultimately targeted by Shinozaki and Matsuyama is the ATT but ours is the ATC. 

In Equation \ref{absolute_dr2}, the G-computation estimator has been augmented with a weighted average of the residuals for the SAT subjects. We shall explain later in this section why such a weighted average is an error-correcting term for the potential bias of the G-computation estimator.\cite{benkeser2017doubly} The corresponding DR augmented IOW estimator for the ATC is: 
\begin{equation}
\widehat{\textrm{ATC}} = 
g 
\underbrace
{
\left (
\frac{1}
{n_0}
\left (
\sum_{i=1}^{n_1}
\hat{w}
_i
\epsilon_i^1
+
\sum_{i=n_1+1}^{n}
\hat{Y_i}^1
\right )
\right )
}
_{\hat{\mu}_0^1}
-
g 
\underbrace
{
\left (
\frac{1}{n_0}\sum_{i=n_1+1}^n Y_i
\right )
}_{\hat{\mu}_0^0},
\label{atc_dr1}
\end{equation}
on the additive scale imposed by link function $g(\cdot)$. As for the non-augmented IOW estimators in Section \ref{subsec32}, we can normalize or stabilize the weights so that they sum to one, thereby ensuring bounded estimates and improving the finite sample properties of the estimator.\cite{dahabreh2020extending} Using the normalized weights, we obtain the \textbf{DR augmented normalized IOW} estimators:
\begin{equation}
\hat{\mu}_0^1
=
\frac{
\sum_{i=1}^{n_1}
\hat{w}
_i
\epsilon_i^1
}
{
\sum_{i=1}^{n_1}
\hat{w}
_i
}
+
\frac{1}{n_0}
\sum_{i=n_1+1}^{n}
\hat{Y_i}^1, 
\label{absolute_dr3}
\end{equation}
\begin{equation}
\widehat{\textrm{ATC}} = 
g 
\underbrace
{
\left (
\frac{
\sum_{i=1}^{n_1}
\hat{w}
_i
\epsilon_i^1
}
{
\sum_{i=1}^{n_1}
\hat{w}
_i
}
+
\frac{1}{n_0}
\sum_{i=n_1+1}^{n}
\hat{Y_i}^1
\right )
}
_{\hat{\mu}_0^1}
-
g 
\underbrace
{
\left (
\frac{1}{n_0}\sum_{i=n_1+1}^n Y_i
\right )
}_{\hat{\mu}_0^0}.
\label{atc_dr2}
\end{equation}
Equations \ref{absolute_dr3} and \ref{atc_dr2} should provide more stable and precise estimation than Equations \ref{absolute_dr2} and \ref{atc_dr1}, respectively.\cite{dahabreh2020extending} 

Our main contribution is combining the entropy balancing-based MAIC approach, described in Section \ref{subsec33}, with the G-computation estimator. Again, based on the outcome model fitted to the SAT, let $\hat{Y_i}^1=q^{-1} \left ( m (\mathbf{X}_i; \hat{\mathbf{\beta}} ) \right )$ be a prediction of the potential outcome $Y_i^1$ under the active intervention for all subjects $i=1,\dots,n$ in the SAT and the external control. We now employ the MAIC weights $\hat{v}_i$ for $i=1,\dots,n_1$ derived in Equation \ref{weight_equation},  instead of the ``modeling'' IOW weights, to estimate the error-correcting term used to augment the G-computation estimator. The resulting \textbf{DR augmented MAIC} estimator of the mean absolute outcome $\mu_0^1$ is: 
\begin{align}
\hat{\mu}_0^1
=&
\sum_{i=1}^{n_1}
\hat{v}
_i
\left (
Y_i
-
\hat{Y}^1_i
\right )
+
\frac{1}{n_0}
\sum_{i=n_1+1}^{n}
\hat{Y_i}^1 \nonumber \\ 
=&
\sum_{i=1}^{n_1}
\hat{v}
_i
\epsilon_i^1
+
\frac{1}{n_0}
\sum_{i=n_1+1}^{n}
\hat{Y_i}^1,
\label{absolute_maicdr}
\end{align}
where the G-computation estimator has been augmented with a weighted average of the residuals $\epsilon_i^1=Y_i -\hat{Y}_i^1$ for $i=1,\dots,n_1$, but this time the weighted average has been computed using the MAIC weights. 

The corresponding DR augmented MAIC estimator for the ATC is: 
\begin{equation}
\widehat{\textrm{ATC}} = 
g 
\underbrace
{
\left (
\sum_{i=1}^{n_1}
\hat{v}
_i
\epsilon_i^1
+
\frac{1}{n_0}
\sum_{i=n_1+1}^{n}
\hat{Y_i}^1
\right )
}
_{\hat{\mu}_0^1}
-
g 
\underbrace
{
\left (
\frac{1}{n_0}\sum_{i=n_1+1}^n Y_i
\right )
}_{\hat{\mu}_0^0},
\label{atc_maicdr}
\end{equation}
on the additive scale imposed by link function $g(\cdot)$. We conjecture that the DR augmented MAIC estimators in Equations \ref{absolute_maicdr} and \ref{atc_maicdr} will perform better statistically than the augmented estimators based on the modeling approach to weighting, which could exhibit erratic performance with highly variable weights, particularly if these are combined with a misspecified outcome model.\cite{robins2007comment, kang2007demystifying} We expect the DR augmented MAIC estimators to inherit the more attractive properties of the entropy balancing weights: (1) lower susceptibility to bias by directly enforcing covariate balance; and (2) greater stability, translating into larger effective sample sizes after weighting and enhanced precision in estimation. 

The augmented weighting estimators in Equations \ref{absolute_dr2}, \ref{absolute_dr3} and \ref{absolute_maicdr} are doubly robust for the mean potential outcome $\mu^1_0$. That is, they estimate $\mu^1_0$ consistently as long as either the propensity score model for data source assignment or the outcome model is correctly specified, but not necessarily both. Note that, contrary to the approach that we shall describe in Section \ref{subsec36}, this double robustness does not depend on using a canonical link function for the outcome model. In the Appendix, we provide an intuitive heuristic to demonstrate double robustness, which illustrates why the weighted average of residuals is an error-correcting term for the potential bias of G-computation. The heuristic is summarized as follows. Consider that all the augmented weighting estimators have the general form:  
\begin{equation}
\hat{\mu}_0^1
=
\sum_{i=1}^{n_1}
\hat{u}_{i}
\left
(
Y_i
-
\hat{Y}^1_i
\right )
+
\frac{1}{n_0}
\sum_{i=n_1+1}^{n}
\hat{Y_i}^1,
\label{general_form}
\end{equation}
for a generic weight estimate $\hat{u}_i$, where $\hat{u}_i=\hat{w}_i/n_0$ for the DR augmented IOW estimator in Equation \ref{absolute_dr2}, $\hat{u}_i=\hat{w}_i/ \sum_{i=1}^{n_1} \hat{w}_i$ for the DR augmented normalized IOW estimator in Equation \ref{absolute_dr3}, and $\hat{u}_i=\hat{v}_i$ for the DR augmented MAIC estimator in Equation \ref{absolute_maicdr}. 

If the outcome model is correctly specified, the expectation of the first summation in Equation \ref{general_form} converges to zero as $n_1 \rightarrow \infty$ because $\hat{Y}_i^1 \rightarrow Y_{i}$ and the terms inside the summation cancel out, irrespective of any postulated propensity score model for data source assignment. The second summation is equivalent to the G-computation estimator and is consistent for $\mu_0^1$ because the outcome model is correct. Consequently, $\hat{\mu}_0^1 \rightarrow \mu_0^1$ and $\widehat{\textrm{ATC}} \rightarrow \textrm{ATC}$, under the assumption that $\hat{\mu}_0^0=\frac{1}{n_0}\sum_{i=n_1+1}^n Y_i$ is consistent for $\mu_0^0$. Conversely, if the propensity score model is correctly specified but the outcome model is incorrect, the first summation consistently cancels out the bias produced by the G-computation estimator in the second summation and the remainder term is exactly equal to the non-augmented weighting estimator, which converges to $\mu_0^1$ as $n_1 \rightarrow \infty$ because the propensity score model is correct. Similarly, $\widehat{\textrm{ATC}} \rightarrow \textrm{ATC}$ (assuming $\hat{\mu}_0^0 \rightarrow \mu_0^0$). Hence, the augmented weighting estimators in Equations \ref{atc_dr1}, \ref{atc_dr2} and \ref{atc_maicdr} are doubly robust for the ATC. 

While all the augmented weighting estimators described in this section are doubly robust, the DR augmented MAIC estimators defined in Equations
\ref{absolute_maicdr} and \ref{atc_maicdr} are arguably more robust to model misspecification bias because they are consistent under a greater number of distinct underlying data-generating mechanisms. Namely, DR augmented MAIC is consistent as long as either: (1) the log-odds of the propensity score are linear across the covariate balance functions; (2) the potential outcome under the active intervention is linear across the covariate balance functions; or (3) the explicit augmentation model for the potential outcome under the active intervention is correctly specified. The estimation of the weights is consistent as long as either the first or the second condition holds. Conversely, the augmented estimators based on the modeling approach to weighting (defined in Equations \ref{absolute_dr2} to \ref{atc_dr2}) are consistent as long as either the first or the third condition holds, with the first condition being necessary for consistent estimation of the weights. 

\subsection{Weighted G-computation}\label{subsec36}

In Section \ref{subsec35}, we proposed augmented estimators that combine the predictions of an unweighted outcome model with weights in a weighted average. Nevertheless, there are other ways of constructing augmented weighting estimators. One approach popularized by Park et al (2024) \cite{park2024unanchored} that has been  claimed to be doubly robust consists of G-computation based on the predictions of a weighted outcome model.\cite{dahabreh2020extending,
park2024unanchored} Where the target estimand is the ATC, this involves: (1) estimating weights using the methods described in Section \ref{subsec32} and Section \ref{subsec33}; (2) fitting a weighted model for the conditional outcome expectation to the SAT participants; and (3) marginalizing the outcome predictions of the weighted regression over the external control covariate distribution. The resulting estimator for the mean absolute outcome $\mu_0^1$ would be: 
\begin{equation}
\hat{\mu}_0^1= 
\frac{1}
{n_0}
\sum_{i=n_1+1}^{n}
\hat{Y}_i^1
=
\frac{1}
{n_0}
\sum_{i=n_1+1}^{n}
q^{-1} \left (m({\mathbf{X}_i};\hat{\mathbf{\beta}}_v)\right), 
\label{weighted_Gcomp_absolute}
\end{equation}
where $m({\mathbf{X}_i};\hat{\mathbf{\beta}}_v)$ indexes the fitted weighted regression with vector $\hat{\mathbf{\beta}}_v$ of parameter estimates. The ATC would be estimated by substituting Equation \ref{weighted_Gcomp_absolute} into Equation \ref{GComp_ATC}. 

Such estimators are only doubly robust where the outcome model is a generalized linear model (GLM) with a canonical link function $q(\cdot)$.\cite{dahabreh2020extending, gabriel2024inverse,robins2007comment, wooldridge2007inverse} The estimator in Equation \ref{weighted_Gcomp_absolute} and the corresponding ATC estimator would not be doubly robust where the GLM link function is non-canonical,\cite{gabriel2024inverse} or where the outcome model is a Cox proportional hazards model or a parametric survival model in the time-to-event setting.\cite{gabriel2024propensity} Nevertheless, results by Gabriel et al (2024) suggest asymptotic equivalence and similar finite sample performance to the augmented weighting estimators in Section \ref{subsec35} for GLMs with canonical link functions fitted via maximum-likelihood,\cite{gabriel2024inverse} provided that the same weights are used and correct model specification. We note that the target of the investigations by Gabriel et al (2024) is the ATE and the modeling approach to weighting.\cite{gabriel2024inverse}

\subsection{Variance estimation}\label{subsec37}

To estimate the variance and construct confidence intervals (CIs) for $\hat{\mu}_{0}^1$ and $\widehat{\textrm{ATC}}$, it is possible to use empirical sandwich-type (``robust'') variance estimators to account for the correlation induced by weighting.\cite{phillippo2016nice, signorovitch2010comparative} In the specific context of non-randomized comparisons, such as the externally controlled SATs and unanchored ITCs explored in this article, these estimators {have exhibited either under-precision or over-precision for the ATT (or the ATC)\cite{reifeis2022variance, kostouraki2024variance} and under-precision for the ATE.\cite{austin2016variance,hernan2000marginal}} This is because most implementations ignore the estimation of the propensity score model or the weights, assuming the weights to be fixed quantities.\cite{reifeis2022variance, kostouraki2024variance} 



Analytic expressions that incorporate weight estimation could be derived,\cite{cheng2020statistical, reifeis2022variance, kostouraki2024variance} but we propose a practical alternative based on the ordinary non-parametric bootstrap,\cite{austin2016variance} explicitly accounting for uncertainty in the weight estimation. This involves resampling with replacement the concatenated IPD consisting of the SAT and external control data. In each bootstrap iteration, the weight estimation and/or outcome modeling procedures are performed, and $\mu_{0}^1$, $\mu_{0}^0$ and $\textrm{ATC}$ are re-estimated.  Standard errors for $\hat{\mu}_{0}^1$, $\hat{\mu}_{0}^0$ and $\widehat{\textrm{ATC}}$,  are given by the standard deviations across the bootstrap resamples. Subsequently, Wald-type CIs can be constructed. Alternatively, one can directly calculate CIs from the percentiles of the bootstrap resamples, e.g., 2.5\% and 97.5\% for the 95\% CI. 

\subsection{External controls with unavailable individual participant data}\label{subsec38}

In the context of unanchored ITCs, the external control is often a historical trial for which IPD are unavailable, due to privacy or confidentiality reasons. In this case, only published AD are available for the external control.\cite{phillippo2016nice, phillippo2018methods, lambert2023enriching} Such data consist of marginal summary moments $\hat{\mathbf{\theta}}$ from reported tables of baseline characteristics, typically only including means and standard deviations (for continuous covariates), and an estimate $\hat{\mu}_0^0$ of the mean absolute outcome under the control in the external data source.\cite{phillippo2016nice, phillippo2018methods, signorovitch2010comparative, wang2021matching, jackson2021alternative} An important shortfall of this scenario is the need to assume that $\hat{\mathbf{\theta}}_j = \mathbf{\theta}_j$ with zero variability for the covariate balance function moments $j=1,\dots,k$, i.e., that the external control covariate distributional data are fixed.\cite{josey2021transporting} While this may be reasonable with large sample sizes for the external control, it can otherwise result in overly precise inferences and inflated Type I error rates.\cite{josey2021transporting}

In this setting, for all methods except (non-augmented) MAIC, one must simulate $M$ individual-level covariate profiles from the assumed covariate distribution of the external control based on published summary statistics.\cite{remiro2022parametric, remiro2024transportability, ren2024advancing} The number $M$ of hypothetical subject profiles should be relatively large, e.g., $M=1000$, to minimize sampling variability and random seed sensitivity, and does not necessarily need to match the original sample size $n_0$ of the external control.\cite{remiro2022parametric, remiro2024transportability, ren2024advancing} Necessary information to infer the joint covariate distribution of the external control, e.g., distributional forms and correlation structures, is rarely published. Hence, this must be borrowed from other data sources or selected based on theoretical properties, following recommendations in the literature.\cite{phillippo2016nice, phillippo2018methods, remiro2022parametric, remiro2024transportability,  ren2024advancing}  

The notation and procedures for Section \ref{subsec32}, Section \ref{subsec34}, Section \ref{subsec35} and Section \ref{subsec36} change as follows. The observed IPD for the SAT is stacked with the simulated subject-level covariate data for the external control. The concatenated dataset is now $(S_i, \mathbf{X}_i, T_i, Y_i)$ for $i=1,\dots, n_1, n_1+1, \dots, n_1+M$. For the SAT subjects $i=1,\dots, n_1$, we have $S_i=1$ and $T_i=1$, with $\mathbf{X}_i$ and $Y_i$ corresponding to the actual covariate and outcome values observed in the trial. For the hypothetical external controls $i=n_1+1,\dots,n_1+M$, we have $S_i=0$ and $T_i=0$, the values of $\mathbf{X}_i$ are simulated, and $Y_i$ are unavailable but not required for the analysis. This is because the target estimand is the ATC and the outcomes for the external group under the control have already been observed factually, with the mean estimate $\hat{\mu}_0^0$ available from published results.

The general form of the IOW estimators for the ATC, described in Section \ref{subsec32}, is now: 
\begin{equation*}
\widehat{\textrm{ATC}} 
= 
g 
\underbrace
{
\left ( 
\frac{1}{K}
\sum_{i=1}^{n_1} \hat{w}_i Y_i
\right )
}
_{\hat{\mu}_0^1}
-
g 
\left (
\hat{\mu}_0^0
\right ),
\end{equation*}
where $K$ is a constant. There is only a change in notation here given that individual-level outcomes under the control are now unavailable for the external data source, and cannot be included in the concatenated dataset. 

For the G-computation estimator for the ATC, outlined in Section \ref{subsec34}, we now have:  
\begin{equation}
\widehat{\textrm{ATC}} 
= 
g 
\underbrace
{
\left (
\frac{1}
{M}
\sum_{i=n_1+1}^{n_1+M}
\hat{Y}_i^1
\right )
}_{\hat{\mu}_0^1}
-
g 
\left (
\hat{\mu}_0^0
\right ),
\label{GComp_ATC_noIPD}
\end{equation}
where the potential outcome predictions $\hat{Y}_i^1$ under the active intervention are generated for each hypothetical external control subject $i=n_1+1,\dots,n_1+M$, and averaged over the simulated covariate profiles. 

The general form of the DR augmented weighting estimators for the ATC, proposed in Section \ref{subsec35}, is now:
\begin{equation*}
\widehat{
\textrm{
ATC}}
=
g
\underbrace{
\left (
\frac{1}{K}
\sum_{i=1}^{n_1}
\hat{\omega}
_i
\epsilon_i^1
+
\frac{1}{M}
\sum_{i=n_1+1}^{n_1+M}
\hat{Y_i}^1 
\right )
}_{\hat{\mu}_0^1}
-
g
\left (
\hat{\mu}_0^0
\right ),
\end{equation*}
where $K$ is a constant, $\epsilon_i^1=Y_i -\hat{Y}_i^1$ and $\hat{\omega}_i$ are a residual term and a weight estimate, respectively, for $i=1,\dots,n_1$, and the potential outcome predictions $\hat{Y}_i^1$ under the active intervention are generated for all SAT subjects and hypothetical external controls $i=1,\dots,n_1,n_1+1,\dots, n_1+M$.   

For the weighted G-computation estimator in Section \ref{subsec36}, the outcome predictions would be averaged over the simulated covariates for the external control. The resulting estimator for the mean absolute outcome $\mu_0^1$ is $\hat{\mu}_0^1= 
\frac{1}
{M}
\sum_{i=n_1+1}^{n_1+M}
\hat{Y}_i^1
=
\frac{1}
{M}
\sum_{i=n_1+1}^{n_1+M}
q^{-1} \left (m({\mathbf{X}_i};\hat{\mathbf{\beta}}_v)\right)$, which would be input into Equation \ref{GComp_ATC_noIPD} for estimation of the ATC.

The unavailability of IPD for the external control entails some changes to the non-parametric bootstrap procedure described in Section \ref{subsec37} for variance estimation. In this case, only the SAT data, $(S_i, \mathbf{X}_i, T_i, Y_i)$ for $i=1,\dots, n_1$, are resampled to re-estimate $g(\mu_0^1)$ in each bootstrap iteration, with the standard error, $\textrm{SE} \left (g(\hat{\mu}^1_0)\right)$, computed as the standard deviation over the bootstrap resamples. Then, the decomposition:
\begin{equation}
\textrm{SE}
\left(\widehat{\textrm{ATC}}\right )
=
\sqrt{\big(\textrm{SE}
\left 
(
g(\hat{\mu}^1_0)
\right
)\big)^2 + \big(\textrm{SE}
\left 
(
g(\hat{\mu}^0_0)
\right
)\big)^2},
\label{variance_limited_IPD}
\end{equation}
is used to estimate the standard error of the ATC, where $\textrm{SE} \left 
(g(\hat{\mu}^0_0)
\right
)$ is derived from published aggregate-level results.\cite{remiro2022parametric, chandler2024uncertain, ren2024advancing} A limitation of the above formula is that it assumes that the mean absolute outcomes are statistically independent. Moreover, while computing $\textrm{SE} \left 
(g(\hat{\mu}^0_0)
\right
)$ is a trivial exercise for continuous and binary outcomes, (e.g., there is a closed-form formula for the standard error of the log-odds using the Delta method), it can be challenging for other outcomes such as those in the time-to-event setting.\cite{chandler2024uncertain, ren2024advancing} Once $\textrm{SE}
\left(\widehat{\textrm{ATC}}\right )$ is computed, Wald-type CIs can be readily constructed.

\section{Simulation study}\label{sec4}

\subsection{Aims}\label{subsec41}

We conducted a simulation study to evaluate the performance of various estimators under different conditions. The simulation study design was planned following the structured ``ADEMP'' approach outlined by Morris et al (2019),\cite{morris2019using} to ensure reproducibility and meaningful conclusions. Specifically, we clearly defined research aims, data-generating mechanisms under controlled scenarios and estimands, and assessed the performance of several estimators using relevant performance measures: bias, empirical standard error and coverage. All simulations and analyses were performed using R statistical software version 4.3.1.\cite{r2017r} \textcolor{black}{The files and code required to run the simulations are publicly available on Github at \url{https://github.com/harlanhappydog/DRAWE-}}.

\subsection{Data-generating mechanisms}\label{subsec42}


We simulated data inspired by the data-generating mechanisms in a simulation study by Kang and Schafer (2007).\cite{kang2007demystifying} Some modifications were required since Kang and Schafer (2007) considered continuous-valued outcomes,\cite{kang2007demystifying} while we consider binary outcomes. The simulated data consist of $(\mathbf{X}_i, \mathbf{Z}_i, T_i, S_i, Y_i)$ for $i=1,\ldots,n$, with the control group fully external such that $S_i=T_i$, with $n_1= \sum_{i=1}^{n}{S_{i}}$, and $n_0 =\sum_{i=1}^{n}{(1-S_{i})}$, as detailed in Section \ref{subsec31}. While $\mathbf{X}_i$ is observed, $\mathbf{Z}_i$ is unobserved. To generate the data, $\mathbf{X}_i$ is distributed as $\mathrm{Normal}(0,I_4)$, for $i$ in 1,...,$n$, and $\mathbf{Z}_i$ is then obtained by applying the following transformations:
\begin{align*}
Z_{i1} &= \textrm{scale}(\exp(X_{i1}/2)), \\
Z_{i2} &= \textrm{scale}(X_{i2}^{2}),\\
Z_{i3} &= \textrm{scale}((X_{i1}X_{i3} + 0.6)^3), \\
Z_{i4} &= \textrm{scale}((X_{i2} + X_{i4} + 20)^2),
\end{align*}
where $\textrm{scale}(\dot)$ indicates normalization such that  $Z_{i1}, Z_{i2}, Z_{i3}$ and $Z_{i4}$ each have mean of $0$ and standard deviation of $1$, i.e.,  $Z = \textrm{scale}(f(X)) = ( f(X) - \textrm{mean}(f(X)))/{\textrm{sd}(f(X))}$.  Note that these transformations are similar to the ones detailed by Kang and Schafer (2007)\cite{kang2007demystifying} but not identical, with changes made to highlight the consequences of model misspecification.

We consider four different scenarios. For each, we generated 10,000 simulated datasets of size $n=200$ and $n=1000$.  Note that in all four scenarios the distribution of $S$ is approximately balanced such that $n_{1} \approx n_{0}$.  The four scenarios are defined as:

\begin{itemize}
    \item KS1: $Y_i$ is generated from a Bernoulli distribution with \\
           $\quad \textrm{Pr}(Y_i=1 \mid \mathbf{X}_i, T_i) = \mathrm{expit}({X}_{1i} - 1.50{X}_{2i}  + 0.5{X}_{3i} - 0.5{X}_{4i} + 1.50T_{i} - 0.50T_{i}{X}_{1i})$, \\     
    where $T_i=S_i$, and $S_{i}$ is generated from a Bernoulli distribution with \\ $\textrm{Pr}(S_i=1 \mid \mathbf{X}_i) = \mathrm{expit}(-X_{i1} + 0.5 X_{i2} - 0.25 X_{i3} -0.5 X_{i4})$.  \\
 The distribution of the covariates is such that overlap between the two groups is relatively high, with overlap proportions of 0.68, 0.85, 0.92, and 0.85 for  $X_{1}$,$X_{2}$,$X_{3}$,and $X_{4}$, respectively (see Figure \ref{fig:overlap1} in the Supplementary Material).
    
       \item KS2: $Y_i$ is generated from a Bernoulli distribution with \\
     $\quad \textrm{Pr}(Y_i=1 \mid \mathbf{Z}_i, T_i)  = \mathrm{expit}({Z}_{1i} - 1.50{Z}_{2i}  + 0.5{Z}_{3i} - 0.5{Z}_{4i} + 1.50T_{i} - 0.50T_{i}{Z}_{1i})$, \\
    where $T_i=S_i$, and $S_{i}$ is generated from a Bernoulli distribution with \\ $\textrm{Pr}(S_i=1 \mid \mathbf{X}_i) = \mathrm{expit}(-X_{i1} + 0.5 X_{i2} - 0.25 X_{i3} -0.5 X_{i4})$.   \\
    The relevant covariate adjustment approaches would fit an outcome model to the observed $\mathbf{X}_i$, as the $\mathbf{Z}_i$ used for the true outcome-generating process are unobserved.  The distribution of the covariates is such that overlap between the two groups is relatively high, with overlap proportions of 0.68, 0.85, 0.92, and 0.85 for  $X_{1}$,$X_{2}$,$X_{3}$,and $X_{4}$, respectively (see Figure \ref{fig:overlap1} in the Supplementary Material).
     
    \item KS3: $Y_i$ is generated from a Bernoulli distribution with \\
     $\quad \textrm{Pr}(Y_i=1 \mid \mathbf{X}_i, T_i) = \mathrm{expit}({X}_{1i} - 1.50{X}_{2i}  + 0.5{X}_{3i} - 0.5{X}_{4i} + 1.50T_{i} - 0.50T_{i}{X}_{1i})$, \\
    where $T_i=S_i$, and $S_{i}$ is generated from a Bernoulli distribution with \\ $\textrm{Pr}(S_i=1 \mid \mathbf{Z}_i) = \mathrm{expit}(-Z_{i1} + 0.5 Z_{i2} - 0.25 Z_{i3} - 0.5 Z_{i4})$.  \\
     The relevant covariate adjustment approaches would balance or fit a propensity score model to the observed $\mathbf{X}_i$, as the $\mathbf{Z}_i$ used for the true data source assignment process are unobserved. The distribution of the covariates is such that overlap between the two groups is relatively high, with overlap proportions of 0.71, 0.84, 0.99, and 0.89 for  $X_{1}$, $X_{2}$, $X_{3}$, and $X_{4}$, respectively (see Figure \ref{fig:overlap3} in the Supplementary Material).

    \item KS4: $Y_i$ is generated from a Bernoulli distribution with \\
     $\quad \textrm{Pr}(Y_i=1 \mid \mathbf{Z}_i, T_i)  = \mathrm{expit}({Z}_{1i} - 1.50{Z}_{2i}  + 0.5{Z}_{3i} - 0.5{Z}_{4i} + 1.50T_{i} - 0.50T_{i}{Z}_{1i})$, \\
     where $T_i=S_i$, and $S_{i}$ is generated from a Bernoulli distribution with \\ $\textrm{Pr}(S_i=1 \mid \mathbf{Z}_i) = \mathrm{expit}(-Z_{i1} + 0.5 Z_{i2} - 0.25 Z_{i3} - 0.5 Z_{i4})$.  \\
     The distribution of the covariates is such that overlap between the two groups is relatively high, with overlap proportions of 0.71, 0.84, 0.99, and 0.89 for  $X_{1}$, $X_{2}$, $X_{3}$, and $X_{4}$, respectively (see Figure \ref{fig:overlap3} in the Supplementary Material).
\end{itemize}
We assume that there is unlimited access to subject-level data for the SAT and the external control, such that $(\mathbf{X}_i, T_i, S_i, Y_i)$ are observed for all $i=1,\ldots,n$.

\subsection{Estimands}\label{subsec43}

The estimand of interest is the ATC, as defined in Section \ref{sec2}. We adopt the logit link function $g(p) =\ln(p/(1-p))$ for marginal outcome probability $p$, such that the ATC is on the marginal log-odds ratio scale. The values of the ATC estimands were calculated numerically, by simulating 10 million binary outcomes using the true data-generating mechanisms outlined in Section \ref{subsec42}. Data-generating mechanisms KS1, KS2, KS3 and KS4 correspond to true ATCs of 1.116, 1.215, 1.068 and 1.181, respectively. 

\subsection{Methods}\label{subsec44}

We compared 16 estimators:

\begin{itemize}

    \item[1.] The na{\"i}ve estimator, which does not perform covariate adjustment:

\begin{equation}
\widehat{\textrm{ATC}} 
= 
g 
\underbrace
{
\left ( 
\frac{1}{n_1}\sum_{i=1}^{n_1} Y_i
\right )
}_{\hat{\mu}_1^1}
-
g 
\underbrace
{
\left (
\frac{1}{n_0}\sum_{i=n_1+1}^n Y_i
\right )
}_{\hat{\mu}_0^0}.
\label{na{\"i}ve}
\end{equation}
    
    \item[2.] The \textbf{IOW} estimator as per Equation \ref{norm_weights_1} (Section \ref{subsec32}), with weights derived using the ``modeling'' approach with a logistic regression data source assignment model as per Equation \ref{logistic_PS}.

    \item[3.] The \textbf{normalized IOW} estimator with normalized weights as per Equation \ref{norm_weights_2} (Section \ref{subsec32}) derived using the ``modeling'' approach with a logistic regression data source assignment model as per Equation \ref{logistic_PS}.
    
    \item[4.] The \textbf{MAIC} (entropy balancing) estimator, as per Equation \ref{norm_weights_3} (Section \ref{subsec33}).
    
    \item[5.] The \textbf{G-computation} estimator, as per Equation \ref{GComp_ATC} (Section \ref{subsec34}) with the outcome model defined with a logistic link function.

 \item[6.] The \textbf{DR augmented IOW} estimator, as per Equation \ref{atc_dr1} (see Section \ref{subsec35}) with the outcome model  defined with a logistic link function.
    
    \item[7.] The \textbf{DR augmented normalized IOW} estimator, as per Equation \ref{atc_dr2} (see Section \ref{subsec35}) with the outcome model  defined with a logistic link function.
    
    \item[8.] The \textbf{DR augmented MAIC} (augmented entropy balancing) estimator, as per Equation \ref{atc_maicdr} (our main contribution, see Section \ref{subsec35}) with the outcome model  defined with a logistic link function.
    
    \item[9.] The \textbf{weighted G-computation (normalized IOW weights)} estimator described in Section \ref{subsec36} with the outcome model defined with a logistic link function, and using the (normalized) ``modeling'' IOW weights.

    \item[10.] The \textbf{weighted G-computation (MAIC weights)} estimator described in Section \ref{subsec36} with the outcome model defined with a logistic link function, but using the MAIC (entropy balancing) weights instead of the ``modeling'' IOW weights. 

    \item[11-16.] Estimators {11-16} are the same as Estimators 5-10 but using a non-canonical Cauchit link function, $q(x)= \textrm{tan}(\pi(x-0.5))$. See Morgan and Smith (1992)\cite{morgan1992note} for an example of model fitting with the Cauchit.
\end{itemize}
To be clear, for Estimators 1-10, both the outcome model and the propensity score model for data source assignment will be correctly specified in KS1.  For Estimators 11-16, the propensity score model for data source assignment will be correctly specified in KS1, but not the outcome model. In KS2, the outcome model will be incorrectly specified for all estimators fitting an outcome model (Estimators 5-16). Finally, in KS3 the propensity score model will be incorrectly specified for all estimators considering a propensity score model, and in KS4 both the outcome model and the propensity score model will be incorrectly specified.

For all estimators, we computed 95\% CIs using the non-parametric bootstrap approach described in Section \ref{subsec37}.  Specifically, we used $B=200$ bootstrap resamples of the concatenated SAT and external control IPD to approximate the standard error of $\widehat{\textrm{ATC}}$ and subsequently constructed Wald-type CIs.

\subsection{Performance measures}\label{subsec45}

To assess the performance of the estimators in our simulation study, we computed several key metrics: bias, empirical standard error (ESE), and 95\% CI coverage. Bias was calculated as the difference between the average of the point estimates across simulations and the true estimand value, providing a measure of systematic error. The ESE was computed as the standard deviation of the point estimates across simulations, reflecting the precision or variability of the different estimators. The 95\% CI coverage was determined as the proportion of simulated datasets in which the constructed 95\% CI contained the true estimand value, evaluating the quality of interval estimation. In addition, we estimated Monte Carlo standard errors (MCSEs) using the formulas provided by Morris et al (2019) to quantify the uncertainty in the performance measures due to using a finite number of simulations.\cite{morris2019using}

\subsection{Results}\label{subsec46}

The complete results for the simulation study are displayed in Tables \ref{tab:SS1_1_results}-\ref{tab:SS1_4_results}.   Note that, for the  $n=200$ simulations, the MCSEs are less than 0.018 for bias, less than 0.013 for  ESE and less than 0.006 for 95\% CI coverage; for the $n=1,000$ simulations, the MCSEs are less than 0.011 for bias, less than 0.008 for  ESE and less than 0.005 for 95\% CI coverage.

%
Under $n=1000$, all methods perform as expected in terms of bias. The na{\"i}ve estimator appears biased in all four scenarios (but in KS3 the bias is very small), the DR estimators appear unbiased when either model is correct (Scenarios KS1, KS2 and KS3), and the singly robust estimators avoid bias only when the corresponding propensity score or outcome model is correct.  Unlike the three DR methods, the weighted G-computation estimators show bias in Scenario KS2 when the Cauchit link function is used in the outcome model (but not when the canonical logistic link is used).  This suggests that the weighted G-computation estimators may be doubly robust for the ATC when the canonical logit link is used, but not otherwise. The trends observed for the bias are similar under $n=200$ with the caveat that the (augmented and non-augmented) weighting estimators that appeared unbiased under $n=1000$, exhibit some small-sample bias in the corresponding scenarios. This is particularly notable in Scenario KS2 and, to a lesser extent, in Scenario KS1, and is probably due to small effective sample sizes after weighting.

Under correct specification of the outcome model (KS1 and KS3), G-computation is the most precise covariate-adjusted estimator, but the augmented estimators are almost as precise (e.g., compare the G-computation estimator which obtains ESE = 0.150 to the DR augmented MAIC estimator which obtains ESE = 0.170 for KS1 with $n=1000$; see Table \ref{tab:SS1_1_results}). Moreover, the augmented estimators generally produce precision gains versus their respective non-augmented weighting counterparts. When both the outcome model and the propensity score model are correctly specified (KS1), all augmented estimators have increased precision compared to the non-augmented weighting estimators based on modeling weights, but not necessarily against MAIC (any increase in precision for $n=1000$ is modest).  When only the propensity score model is correctly specified (KS2), outcome model misspecification does not induce any meaningful loss in precision for the augmented estimators compared to their non-augmented weighting counterparts.

 There have been some concerns in the literature about doubly robust augmented estimators amplifying bias and variance when misspecified weights are combined with a misspecified outcome model.\cite{kang2007demystifying} Such amplification is not apparent in our simulation study. In KS4, our proposed DR augmented MAIC estimator is the least biased of all estimators and the most precise of the augmented and non-augmented weighting estimators.  Under $n=1000$, this advantage is modest when the logit link is used for the outcome model (bias of 0.482 versus 0.512 for G-computation), but more pronounced when the Cauchit link is used (bias of 0.350 versus 0.524 for G-computation).

\clearpage

\begin{table}[!b]
    \centering
    \begin{tabular}{|p{0.55\linewidth}|c|c|p{0.075\linewidth}|p{0.075\linewidth}|}
        \hline
         \textbf{Method} &  \textbf{Bias} & \textbf{ESE} & \textbf{95\% CI coverage} & \textbf{Average 95\% CI width} \\
\hline
  \multicolumn{5}{|l|}{$n=200$, $\textrm{ATC} = 1.116$} \\  
  \hline
1. The naive estimator & 0.623 & 0.326 & 0.539 & 1.295 \\ 
  2. IOW & 0.027 & 0.516 & 0.944 & 1.979 \\ 
  3. Normalized IOW & 0.055 & 0.456 & 0.945 & 1.761 \\ 
  4. MAIC & 0.037 & 0.413 & 0.962 & 2.276 \\ 
  \textbf{With logit link used for outcome model:} &&&&\\
  5. G-computation & 0.019 & 0.351 & 0.956 & 1.431 \\ 
  6. DR augmented IOW & 0.029 & 0.424 & 0.955 & 1.663 \\ 
  7. DR augmented normalized IOW & 0.029 & 0.403 & 0.949 & 1.605 \\ 
  8. DR augmented MAIC & 0.031 & 0.406 & 0.955 & 1.718 \\ 
  9. Weighted G-computation (normalized IOW weights) & 0.029 & 0.401 & 0.942 & 1.582 \\ 
  10. Weighted G-computation (MAIC weights) & 0.026 & 0.402 & 0.946 & 1.770 \\ 
  \textbf{With Cauchit link used for outcome model:} &&&&\\
  5. G-computation & 0.006 & 0.383 & 0.965 & 1.689 \\ 
  6. DR augmented IOW & 0.026 & 0.432 & 0.954 & 1.682 \\ 
  7. DR augmented normalized IOW & 0.026 & 0.413 & 0.951 & 1.635 \\ 
  8. DR augmented MAIC & 0.028 & 0.408 & 0.957 & 1.717 \\ 
  9. Weighted G-computation (normalized IOW weights) & 0.063 & 0.914 & 0.964 & 2.696 \\ 
  10. Weighted G-computation (MAIC weights) & 0.052 & 0.415 & 0.963 & 1.857 \\ 
   \hline
  \multicolumn{5}{|l|}{$n=1000$, $\textrm{ATC} = 1.116$} \\
   \hline
1. The naive estimator & 0.604 & 0.142 & 0.008 & 0.562 \\ 
  2. IOW & 0.008 & 0.205 & 0.949 & 0.810 \\ 
  3. Normalized IOW & 0.012 & 0.197 & 0.942 & 0.750 \\ 
  4. MAIC & 0.008 & 0.172 & 0.944 & 0.660 \\ 
    \textbf{With logit link used for outcome model:} &&&&\\
  5. G-computation & 0.005 & 0.150 & 0.950 & 0.593 \\ 
  6. DR augmented IOW & 0.007 & 0.175 & 0.948 & 0.679 \\ 
  7. DR augmented normalized IOW & 0.006 & 0.173 & 0.947 & 0.669 \\ 
  8. DR augmented MAIC & 0.007 & 0.170 & 0.942 & 0.653 \\ 
  9. Weighted G-computation (normalized IOW weights) & 0.006 & 0.170 & 0.940 & 0.648 \\ 
  10. Weighted G-computation (MAIC weights) & 0.007 & 0.171 & 0.939 & 0.649 \\ 
   \textbf{With Cauchit link used for outcome model:} &&&&\\
  5. G-computation & -0.020 & 0.158 & 0.947 & 0.641 \\ 
  6. DR augmented IOW & 0.006 & 0.178 & 0.946 & 0.683 \\ 
  7. DR augmented normalized IOW & 0.006 & 0.175 & 0.945 & 0.673 \\ 
  8. DR augmented MAIC & 0.006 & 0.172 & 0.940 & 0.658 \\ 
  9. Weighted G-computation (normalized IOW weights) & -0.003 & 0.180 & 0.948 & 0.781 \\ 
  10. Weighted G-computation (MAIC weights) & -0.004 & 0.173 & 0.942 & 0.677 \\ 
\hline
    \end{tabular}
    \caption{Results from Scenario KS1, where both the logit-link outcome model and the propensity score model are correctly specified.}
    \label{tab:SS1_1_results}
\end{table}

\clearpage

\begin{table}[!b]
\centering
\begin{tabular}{|p{0.55\linewidth}|c|c|p{0.075\linewidth}|p{0.075\linewidth}|}
        \hline
        \textbf{Method} &  \textbf{Bias} & \textbf{ESE} & \textbf{95\% CI coverage} & \textbf{Average 95\% CI width} \\ 
\hline
  \multicolumn{5}{|l|}{$n=200$, $\textrm{ATC} = 1.215$} \\
 \hline
1. The naive estimator & 0.230 & 0.324 & 0.909 & 1.279 \\ 
  2. IOW & 0.023 & 0.642 & 0.929 & 2.393 \\ 
  3. Normalized IOW & 0.060 & 0.511 & 0.940 & 1.952 \\ 
  4. MAIC & 0.061 & 0.517 & 0.965 & 2.778 \\ 
  \textbf{With logit link used for outcome model:} &&&&\\
  5. G-computation & 0.088 & 0.434 & 0.952 & 1.739 \\ 
  6. DR augmented IOW & 0.050 & 0.532 & 0.949 & 2.093 \\ 
  7. DR augmented normalized IOW & 0.053 & 0.526 & 0.942 & 2.007 \\ 
  8. DR augmented MAIC & 0.048 & 0.489 & 0.953 & 2.042 \\ 
  9. Weighted G-computation (normalized IOW weights) & 0.056 & 0.480 & 0.937 & 1.868 \\ 
  10. Weighted G-computation (MAIC weights) & 0.036 & 0.480 & 0.949 & 2.241 \\ 
 \textbf{With Cauchit link used for outcome model:} &&&&\\
  5. G-computation & 0.222 & 0.506 & 0.951 & 2.085 \\ 
  6. DR augmented IOW & 0.046 & 0.517 & 0.955 & 2.053 \\ 
  7. DR augmented normalized IOW & 0.046 & 0.505 & 0.950 & 1.974 \\ 
  8. DR augmented MAIC & 0.039 & 0.474 & 0.958 & 1.990 \\ 
  9. Weighted G-computation (normalized IOW weights) & 0.225 & 1.950 & 0.964 & 5.768 \\ 
  10. Weighted G-computation (MAIC weights) & 0.132 & 0.486 & 0.972 & 2.478 \\ 
   \hline
  \multicolumn{5}{|l|}{$n=1000$, $\textrm{ATC} = 1.215$} \\
  \hline
1. The naive estimator & 0.226 & 0.141 & 0.658 & 0.557 \\ 
2. IOW & 0.020 & 0.283 & 0.950 & 1.085 \\ 
3. Normalized IOW & 0.016 & 0.220 & 0.939 & 0.834 \\ 
4. MAIC & 0.015 & 0.205 & 0.936 & 0.779 \\ 
  \textbf{With logit link used for outcome model:} &&&&\\
5. G-computation & 0.075 & 0.188 & 0.937 & 0.736 \\ 
6. DR augmented IOW & 0.016 & 0.228 & 0.942 & 0.866 \\ 
7. DR augmented normalized IOW & 0.016 & 0.224 & 0.938 & 0.849 \\ 
8. DR augmented MAIC & 0.014 & 0.205 & 0.935 & 0.776 \\ 
9. Weighted G-computation (normalized IOW weights) & 0.017 & 0.203 & 0.936 & 0.778 \\ 
10. Weighted G-computation (MAIC weights) & 0.014 & 0.203 & 0.936 & 0.770 \\ 
 \textbf{With Cauchit link used for outcome model:} &&&&\\
  5. G-computation & 0.228 & 0.284 & 0.817 & 1.012 \\ 
  6. DR augmented IOW & 0.016 & 0.224 & 0.942 & 0.854 \\ 
  7. DR augmented normalized IOW & 0.016 & 0.220 & 0.940 & 0.836 \\ 
  8. DR augmented MAIC & 0.012 & 0.199 & 0.937 & 0.756 \\ 
  9. Weighted G-computation (normalized IOW weights) & 0.163 & 1.270 & 0.935 & 2.820 \\ 
  10. Weighted G-computation (MAIC weights) & 0.114 & 0.189 & 0.905 & 0.743 \\ 
    \hline
\end{tabular}
\caption{Results from Scenario KS2, where the outcome model is incorrectly specified and the propensity score model is correctly specified.}
\label{tab:SS1_2_results}
\end{table}

\begin{table}[!b]
\centering
\begin{tabular}{|p{0.55\linewidth}|c|c|p{0.075\linewidth}|p{0.075\linewidth}|}
        \hline
        \textbf{Method} &  \textbf{Bias} & \textbf{ESE} & \textbf{95\% CI coverage} & \textbf{Average 95\% CI width} \\ 
\hline
  \multicolumn{5}{|l|}{$n=200$, $\textrm{ATC} = 1.068$} \\
  \hline
1. The naive estimator & -0.032 & 0.307 & 0.950 & 1.211 \\ 
  2. IOW & 0.147 & 0.583 & 0.958 & 2.246 \\ 
  3. Normalized IOW & -0.021 & 0.397 & 0.950 & 1.539 \\ 
  4. MAIC & 0.132 & 0.353 & 0.955 & 1.561 \\ 
  \textbf{With logit link used for outcome model:} &&&&\\
  5. G-computation & 0.015 & 0.286 & 0.962 & 1.180 \\ 
  6. DR augmented IOW & 0.028 & 0.344 & 0.964 & 1.404 \\ 
  7. DR augmented normalized IOW & 0.027 & 0.335 & 0.958 & 1.338 \\ 
  8. DR augmented MAIC & 0.024 & 0.313 & 0.962 & 1.296 \\ 
  9. Weighted G-computation (normalized IOW weights) & 0.020 & 0.305 & 0.958 & 1.245 \\ 
  10. Weighted G-computation (MAIC weights) & 0.018 & 0.305 & 0.959 & 1.300 \\   
    \textbf{With Cauchit link used for outcome model:} &&&&\\
  11. G-computation & -0.031 & 0.296 & 0.969 & 1.316 \\ 
  12. DR augmented IOW & 0.008 & 0.351 & 0.964 & 1.412 \\ 
  13. DR augmented normalized IOW & 0.005 & 0.332 & 0.960 & 1.348 \\ 
  14. DR augmented MAIC & 0.010 & 0.316 & 0.965 & 1.323 \\ 
  15. Weighted G-computation (normalized IOW weights) & 0.018 & 0.701 & 0.972 & 2.176 \\ 
  16. Weighted G-computation (MAIC weights) & -0.001 & 0.320 & 0.970 & 1.448 \\ 
   \hline
  \multicolumn{5}{|l|}{$n=1000$, $\textrm{ATC} = 1.068$} \\
  \hline
1. The naive estimator & -0.039 & 0.135 & 0.937 & 0.530 \\ 
  2. IOW & 0.118 & 0.231 & 0.969 & 0.917 \\ 
  3. Normalized IOW & -0.044 & 0.168 & 0.933 & 0.645 \\ 
  4. MAIC & 0.104 & 0.148 & 0.894 & 0.574 \\ 
    \textbf{With logit link used for outcome model:} &&&&\\  
  5. G-computation & 0.005 & 0.124 & 0.949 & 0.489 \\ 
  6. DR augmented IOW & 0.007 & 0.144 & 0.945 & 0.558 \\ 
  7. DR augmented normalized IOW & 0.007 & 0.142 & 0.943 & 0.547 \\ 
  8. DR augmented MAIC & 0.006 & 0.133 & 0.943 & 0.517 \\ 
  9. Weighted G-computation (normalized IOW weights) & 0.006 & 0.130 & 0.943 & 0.508 \\ 
  10. Weighted G-computation (MAIC weights) & 0.006 & 0.130 & 0.944 & 0.506 \\  
  \textbf{With Cauchit link used for outcome model:} &&&&\\
  11. G-computation & -0.053 & 0.124 & 0.926 & 0.496 \\ 
  12. DR augmented IOW & -0.015 & 0.144 & 0.942 & 0.557 \\ 
  13. DR augmented normalized IOW & -0.017 & 0.141 & 0.939 & 0.546 \\ 
  14. DR augmented MAIC & -0.010 & 0.134 & 0.941 & 0.520 \\ 
  15. Weighted G-computation (normalized IOW weights) & -0.044 & 0.131 & 0.937 & 0.606 \\ 
  16. Weighted G-computation (MAIC weights) & -0.045 & 0.130 & 0.933 & 0.518 \\ 
   \hline
\end{tabular}
\caption{Results from Scenario KS3, where the propensity score model is incorrectly specified. The logit-link outcome model is correctly specified; however, when the Cauchit link is used for the outcome model, both models are incorrectly specified.}
\label{tab:SS1_3_results}
\end{table}

\clearpage

\begin{table}[!b]
\centering
\begin{tabular}{|p{0.55\linewidth}|c|c|p{0.075\linewidth}|p{0.075\linewidth}|}
        \hline
        \textbf{Method} &  \textbf{Bias} & \textbf{ESE} & \textbf{95\% CI coverage} & \textbf{Average 95\% CI width} \\ 
  \hline
  \multicolumn{5}{|l|}{$n=200$, $\textrm{ATC} = 1.181$} \\
  \hline
1. The naive estimator & 0.521 & 0.334 & 0.685 & 1.330 \\ 
  2. IOW & 0.807 & 0.778 & 0.912 & 2.793 \\ 
  3. Normalized IOW & 0.582 & 0.470 & 0.760 & 1.840 \\ 
  4. MAIC & 0.613 & 0.463 & 0.790 & 2.032 \\ 
  \textbf{With logit link used for outcome model:} &&&&\\
  5. G-computation & 0.536 & 0.376 & 0.750 & 1.551 \\ 
  6. DR augmented IOW & 0.586 & 0.464 & 0.767 & 1.846 \\ 
  7. DR augmented normalized IOW & 0.580 & 0.446 & 0.753 & 1.770 \\ 
  8. DR augmented MAIC & 0.517 & 0.407 & 0.779 & 1.671 \\ 
  9. Weighted G-computation (normalized IOW weights) & 0.546 & 0.415 & 0.760 & 1.692 \\ 
  10. Weighted G-computation (MAIC weights) & 0.545 & 0.423 & 0.786 & 1.808 \\
   \textbf{With Cauchit link used for outcome model:} &&&&\\
   5. G-computation & 0.555 & 0.363 & 0.799 & 1.677 \\ 
  6. DR augmented IOW & 0.455 & 0.430 & 0.850 & 1.767 \\ 
  7. DR augmented normalized IOW & 0.456 & 0.415 & 0.837 & 1.704 \\ 
  8. DR augmented MAIC & 0.400 & 0.386 & 0.878 & 1.654 \\ 
  9. Weighted G-computation (normalized IOW weights) & 0.575 & 1.291 & 0.871 & 4.635 \\ 
  10. Weighted G-computation (MAIC weights) & 0.515 & 0.405 & 0.882 & 2.135 \\ 
   \hline
  \multicolumn{5}{|l|}{$n=1000$, $\textrm{ATC} = 1.181$} \\
  \hline
1. The naive estimator & 0.495 & 0.146 & 0.069 & 0.575 \\ 
  2. IOW & 0.778 & 0.350 & 0.340 & 1.420 \\ 
  3. Normalized IOW & 0.514 & 0.197 & 0.256 & 0.768 \\ 
  4. MAIC & 0.546 & 0.190 & 0.174 & 0.739 \\ 
    \textbf{With logit link used for outcome model:} &&&&\\
  5. G-computation & 0.512 & 0.161 & 0.110 & 0.636 \\ 
  6. DR augmented IOW & 0.536 & 0.185 & 0.183 & 0.729 \\ 
  7. DR augmented normalized IOW & 0.534 & 0.183 & 0.176 & 0.716 \\ 
  8. DR augmented MAIC & 0.482 & 0.170 & 0.196 & 0.667 \\ 
  9. Weighted G-computation (normalized IOW weights) & 0.524 & 0.179 & 0.159 & 0.714 \\ 
  10. Weighted G-computation (MAIC weights) & 0.533 & 0.181 & 0.162 & 0.708 \\ 
  \textbf{With Cauchit link used for outcome model:} &&&&\\
  5. G-computation & 0.524 & 0.144 & 0.050 & 0.578 \\ 
  6. DR augmented IOW & 0.395 & 0.173 & 0.368 & 0.679 \\ 
  7. DR augmented normalized IOW & 0.400 & 0.171 & 0.347 & 0.666 \\ 
  8. DR augmented MAIC & 0.350 & 0.158 & 0.403 & 0.620 \\ 
  9. Weighted G-computation (normalized IOW weights) & 0.488 & 0.165 & 0.226 & 1.089 \\ 
  10. Weighted G-computation (MAIC weights) & 0.479 & 0.158 & 0.155 & 0.646 \\ 
   \hline
\end{tabular}
\caption{Results from Scenario KS4, where both the outcome model and the propensity score model are incorrectly specified.}
\label{tab:SS1_4_results}
\end{table}

\clearpage

In Section \ref{subsec33}, we hypothesized that entropy balancing weights, like those employed by MAIC, can lead to more stable and precise ATC estimation than inverse odds modeling weights. This appears to be confirmed for the non-augmented estimators in our simulation study; MAIC exhibits greater precision than the approaches using (normalized or non-normalized) IOW modeling weights in all simulation scenarios. Additionally, the precision gains have been inherited by the augmented approaches. For the methods highlighted in Section \ref{subsec35}, estimators using MAIC weights display enhanced precision compared to those using IOW modeling weights in all simulation scenarios, while producing similar levels of bias, even lower bias under dual model misspecification. 

Assuming unbiasedness, interval estimation is appropriate if the coverage is approximately equal to 0.95; poor coverage can arise due to bias or to inadequate variance/interval estimation. Coverage is generally close to 0.95 for all covariate adjustment methods in the simulation scenarios under which they are unbiased, which suggests that our proposed non-parametric bootstrap approach for variance estimation is adequate. In the cases in which the DR methods are unbiased (i.e., KS1, KS2 and KS3), coverage rates are between 0.936 and 0.947 across $n=1000$ scenarios. Interestingly, coverage rates seem to increase for these methods under $n=200$ despite the small-sample bias, lying between 0.942 and 0.964.  Note that, due to computational limitations, the non-parametric bootstrap approach in the simulation study was conducted with only $B=200$ resamples. This may have impacted the observed coverage rates and we suspect that coverage might be more appropriate when using a larger number of resamples.

While some covariate adjustment methods display bias-induced undercoverage in the scenarios under which they are biased (e.g., MAIC in KS3 under $n=1000$ or all estimators in KS4), they may also display adequate coverage (e.g., normalized IOW with $n=200$) because of excessively large standard errors, probably due to low effective sample sizes after weighting. As observed for KS4, bias-induced undercoverage tends to worsen with higher sample sizes, as interval estimates around the wrong target value become narrower. The na{\"i}ve estimator displays discernible undercoverage in KS1, KS2 and KS4 (particularly under $n=1000$), not only due to bias but also due to overprecise standard errors that do not account for covariate differences.

\section{Applied example}\label{sec5}

We now demonstrate the application of some of the methods outlined in Section \ref{sec3} to synthetic lung cancer clinical trial data. The data were obtained from the ``MAIC'' R package, implemented by researchers from the pharmaceutical industry.\cite{iain_t_bennett_roche_2022_6624152}  Our objective is to compare the objective response, a binary outcome $Y$, under two treatments: the active ``intervention'' ($T=1$) and the external ``control’' ($T=0$). The data consist of IPD from a SAT ($S=1$) with $n_1=500$ subjects under the ``intervention'', and AD from an external historical SAT ($S=0$) of $n_0=300$, which makes up the ``control''. The unavailability of IPD for the external control allows us to illustrate the methodological extensions described in Section \ref{subsec38}. R code to reproduce the applied example is provided in the Supplementary Material. 

The target estimand is the ATC on the marginal log-odds ratio scale. Four baseline covariates, one continuous -- age -- and three binary -- sex, the Eastern Cooperative Oncology Group (ECOG) performance status and smoking status --- have been identified as prognostic factors under the intervention, and are imbalanced between the intervention SAT and external control samples. There are no missing values for baseline characteristics and outcomes. Subjects in the intervention SAT are, on average, somewhat older, less likely to be male, more likely to be physically restricted (as indicated by ECOG performance status), and more likely to be smokers, relative to subjects in the external control (Table \ref{tab:roche1}). In addition, the age of subjects in the intervention SAT has substantially greater variance than that of subjects in the external control. 

\begin{table}[!htb]
\centering
\begin{tabular}{lcccc}
  \hline
 \textbf{Covariate} & \textbf{Intervention SAT} & \textbf{External control} &
         \textbf{Normalized IOW} &  \textbf{MAIC}\\         
         & &  & \textbf{weighted SAT} & \textbf{weighted SAT} \\
         & ($n_1=500$) & ($n_0=300$) & ($\textrm{ESS}=153.42$) & ($\textrm{ESS}=157.07$)\\
  \hline
Age in years (mean; SD) & 59.85; 9.01 & 50.06; 3.24 & 49.53; 3.18 & 50.06; 3.24 \\ 
  Sex (proportion male) & 0.38 & 0.49 & 0.45 & 0.49 \\ 
  ECOG (proportion status 1) & 0.41 & 0.35 & 0.30 & 0.35 \\ 
  Smoking (proportion smokers) & 0.32 & 0.19 & 0.20 & 0.19 \\ 
   \hline
\end{tabular}
\caption{Summary statistics of the four baseline covariates identified as imbalanced prognostic factors, before and after weighting using MAIC (entropy balancing) and normalized inverse odds weighting (IOW). The standard deviation of age in the weighted columns is $\sqrt{\sum_{i=1}^{n_{1}} v_{i} (X_{1,i} - \sum_{i=1}^{n_{1}} v_{i} X_{1,i})^2}$, where $X_{1,i}$ and $v_i$ are the age and the weight, respectively, for subject $i=1,\dots, n_1$ in the intervention SAT.} 
\label{tab:roche1}
\end{table}

We consider the na{\"i}ve estimator first, which does not perform covariate adjustment (Equation \ref{na{\"i}ve}). In the intervention SAT, 390 of $n_1=500$ subjects attained objective response, which equates to a 78\% response rate, $\hat{\mu}_1^1=0.78$. This implies a log-odds of response of $g(\hat{\mu}_1^1)=1.266$, where $g(\cdot)=\logit(\cdot)$. In the external control, 120 of $n_0=300$ subjects attained objective response, which equates to a 40\% response rate, $\hat{\mu}_0^0=0.40$. This implies a log-odds of response of $g(\hat{\mu}_0^0)=-0.405$. A na{\"i}ve estimate is obtained by simple subtraction: $\widehat{\textrm{ATC}}_{naive} 
= 1.266 - (-0.405) = 1.671$ and using the Delta method, we obtain: $\textrm{SE}\left (g(\hat{\mu}_0^0)\right) = 0.118$ and $\textrm{SE}\left (g(\hat{\mu}_1^1)\right ) = 0.108$, and a Wald-type 95\% CI of $(1.358, 1.984)$.

Next, consider the \textbf{normalized IOW} estimator. A logistic regression model for the probability of SAT participation, conditional on age, sex, ECOG performance status, smoking status, and age-squared is fitted to the concatenated individual-level data comprising the intervention SAT and $M=10000$ simulated covariate profiles for the external control. The estimated propensity scores are used to derive inverse odds weights, which are subsequently normalized to sum to one, yielding the Hajek-type estimator in Equation \ref{norm_weights_2}. Table \ref{tab:roche1} shows that the IOW-weighted covariate means are close to, but do not exactly match, those of the external control. The effective sample size of the IOW-weighted intervention SAT is 153.42 and standardized mean differences are substantially reduced after weighting (e.g., from $1.445$ to $-0.079$ for age), indicating that the weights achieve adequate, though not exact, balance. Using the normalized IOW weights, the ATC is estimated as: $\widehat{\textrm{ATC}}_{IOW} = 1.333$. Having previously calculated $\textrm{SE}\left(g(\hat{\mu}_0^0)\right) = 0.118$, we use the non-parametric bootstrap with $B=10000$ resamples to obtain $\textrm{SE}\left(g(\hat{\mu}_0^1)\right) = 0.183$, and per Equation \ref{variance_limited_IPD}, we then obtain $\textrm{SE}\left(\widehat{\textrm{ATC}}_{IOW}\right) = 0.216$ and a Wald-type 95\% CI of $(0.911, 1.756)$.

We now consider the \textbf{MAIC} estimator. The positivity assumption is assessed using a method proposed by Glimm and Yau (2022), which verifies whether covariate AD from the external control lie within the convex hull of the SAT covariate space, and whether the MAIC numerical optimization algorithm can converge.\cite{glimm2022geometric} This method is implemented using the ``maicLP'' R function in the ``maicChecks'' R package,\cite{maicchecks} which confirms that a feasible weighting solution to the MAIC convex optimization problem exists, i.e., that there is a set of positive weights that can enforce covariate balance between the intervention SAT and the external control, and that the MAIC numerical optimization algorithm can converge. 

\begin{figure}[h]
    \centering
\includegraphics[width=0.75\linewidth]{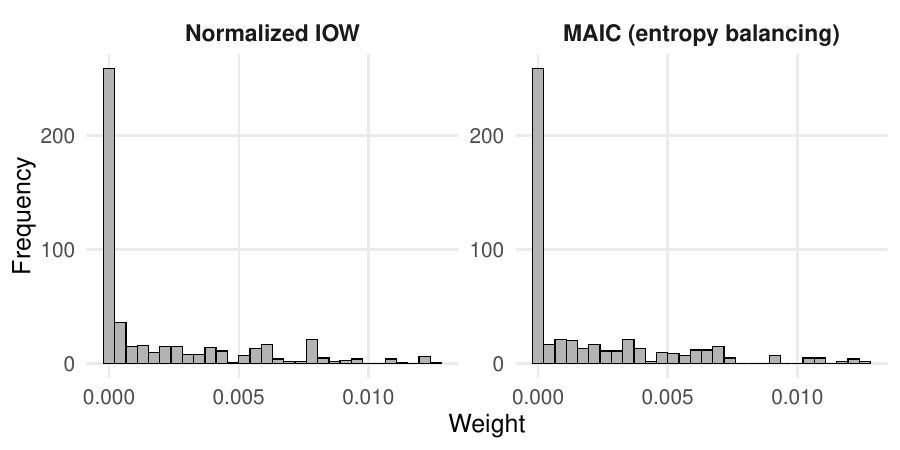}
    \caption{Histogram of the normalized IOW weights (left) and MAIC (entropy balancing) weights (right).}
    \label{fig:roche_hist}
\end{figure}

MAIC is performed using the procedure described in Section \ref{subsec33}. We choose to weight the intervention SAT such that the means of all four baseline covariates and the variance of age are exactly balanced with respect to the external control. Following the notation in Section \ref{subsec33}, we have: 
\begin{align*}
\mathbf{c(X)} = [\mathbf{Age},\mathbf{Sex}, \mathbf{ECOG}, \mathbf{Smoking}, \mathbf{Age}^{2}]^\top = 
\begin{pmatrix}
    45       & 71 & \cdots & 58 \\
   1 & 1 & \cdots & 0 \\
    0 & 0 & \cdots & 1 \\
   0       & 0 & \cdots & 1 \\
   2025 & 5041 & \cdots &  3364
\end{pmatrix},
\end{align*}

\noindent where $\mathbf{c(X)}$ is a 5-by-500 matrix, with the rows representing the age, sex, ECOG performance status, smoking status, and age-squared for subjects in the intervention SAT. Using the BFGS convex optimization algorithm to minimize the objective function in Equation \ref{objective_function}, we obtain $\mathbf{\hat{\gamma}} = (3.542, 0.589, -0.698, -0.048, -0.036)$, and weights are calculated subject to the constraint that they sum to one. Figure \ref{fig:roche_hist} shows a histogram illustrating the empirical distribution of the resulting MAIC weights compared to the normalized IOW weights. The effective sample size (ESS) -- that is, the number of independent non-weighted observations that would be required to give an estimate with approximately the same precision as the weighted sample estimate -- of the intervention SAT after weighting is 157.07. Using the MAIC weights, the ATC is estimated as: $\widehat{\textrm{ATC}}_{MAIC} = 1.331$.  Having previously calculated $\textrm{SE}\left (g(\hat{\mu}_0^0)\right) = 0.118$, we use the non-parametric bootstrap with $B=10000$ resamples to obtain $\textrm{SE}\left(g(\hat{\mu}_0^1)\right)=0.177$, and per Equation \ref{variance_limited_IPD}, we then obtain $\textrm{SE}\left(\widehat{\textrm{ATC}}_{MAIC}\right) = 0.212$ and a Wald-type 95\% CI of $(0.915, 1.748)$.

To perform \textbf{G-computation} with unavailable IPD for the external control, we first simulate $M=10000$ individual-level covariate profiles from an assumed joint covariate distribution for the external control, as per the approach outlined by Remiro-Az\'ocar et al (2022)\cite{remiro2022parametric} or the ``infinite population'' STC method described by Zhang et al (2024).\cite{zhang2024four} We proceed by assuming that the pairwise correlation structure of the four covariates in the external control is equal to that observed in the intervention SAT: 
\begin{align*}    
\begin{pmatrix}
1.00 & 0.03 & 0.00 & 0.00 \\
0.03 & 1.00 & -0.14 & -0.02 \\
 0.00 & -0.14 & 1.00 & -0.01 \\
 0.00 & -0.02 & -0.01 & 1.00 \\
\end{pmatrix},
\end{align*}
with the rows/columns in the order: age, sex, ECOG performance status and smoking status. For age, we assumed a Normal(50.06, 3.24) marginal distribution; for sex, ECOG performance status and smoking status, we assumed Bernoulli(0.49), Bernoulli(0.35) and Bernoulli(0.19) marginal distributions, respectively, based on the summary statistics of the external control in Table \ref{tab:roche1}. Individual-level covariates were ultimately simulated from a Gaussian copula characterized by the aforementioned marginal distributions and pairwise correlation structure, using the ``add\_integration'' function from the multinma R package.\cite{phillippo2024multinma}

Subsequently, a logistic-link binomial GLM for the outcome expectation under the intervention, conditional on baseline covariates, was postulated. This relates objective response $Y_i^1$ under the intervention $T=1$ to baseline covariates $\mathbf{X}_i=\left(Age_i, Sex_i, ECOG_i, Smoking_i\right)^\top$ as: 
\begin{equation*}
\textrm{logit}\left(
\textrm{E}(Y_i^1 \mid \mathbf{X}_i; \mathbf{\beta})
\right) = \beta_{0} + \beta_{1}Age_{i} + \beta_{2}Sex_{i} +\beta_{3}ECOG_{i} + \beta_{4}Smoking_{i} + \beta_{5}Age_{i}^{2},
\label{outcome_model_applied}
\end{equation*}
for $i$ in 1,...,$n_{1}$, where $\mathbf{\beta}=\left(\beta_{0},\beta_{1},\beta_{2}, \beta_{3}, \beta_{4}, \beta_{5}\right)^\top$ are regression coefficients. The model was fitted to the intervention SAT using maximum-likelihood estimation with Fisher scoring, and we obtained regression coefficient estimates of $\hat{\beta}_{0} =5.72$, $\hat{\beta}_{1} =-0.20$, $\hat{\beta}_{2} =0.12$, $\hat{\beta}_{3} =0.13$, $\hat{\beta}_{4} =0.01$, and $\hat{\beta}_{5} =0.00$.  Then, following the G-computation procedure outlined in Section \ref{subsec34}, the ATC is estimated as: $\widehat{\textrm{ATC}}_{Gcomp} = 1.325$.  Using the non-parametric bootstrap with $B=10000$ resamples, we obtain $\textrm{SE}\left(g(\hat{\mu}_0^1)\right)=0.164$, and, as per Equation \ref{variance_limited_IPD}, then obtain $\textrm{SE}\left(\widehat{\textrm{ATC}}_{Gcomp}\right) =0.202$, and a Wald-type 95\% CI of $(0.929, 1.722)$.

Finally, our \textbf{DR augmented MAIC} estimator proposed in Section \ref{subsec35} produces the estimate: ${\textrm{ATC}}_{DR}= 1.332$.  Using the non-parametric bootstrap with $B=10000$ resamples, we obtain $\textrm{SE}\left(g(\hat{\mu}_0^1)\right)=0.179$. Then, as per Equation \ref{variance_limited_IPD}, we have $\textrm{SE}\left(\widehat{\textrm{ATC}}_{DR}\right) = 0.214$ and a Wald-type 95\% CI of $(0.912, 1.751)$.

\begin{figure}[!t]
    \centering
    \includegraphics[width=0.7\linewidth]{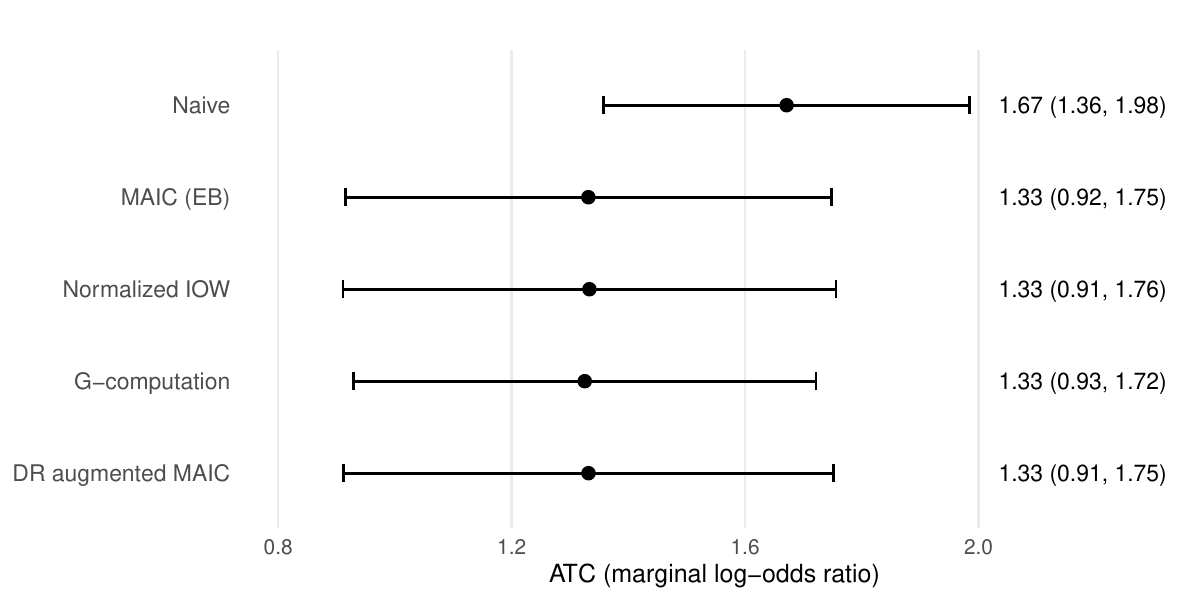}
    \caption{Point estimates with 95\% CIs of the ATC (marginal log-odds ratio of objective response) for the different estimators in the applied example. DR denotes doubly robust and EB denotes entropy balancing.}
    \label{fig:roche}
\end{figure}

Figure \ref{fig:roche} shows the point estimates obtained using the five different estimators alongside their 95\% CIs. When comparing the covariate-adjusted approaches to the na{\"i}ve approach, we observe that covariate adjustment shifts the point estimate towards the null considerably.  When comparing \textbf{normalized IOW}, \textbf{MAIC}, \textbf{G-computation} and our proposed \textbf{DR augmented MAIC} estimator, results across the four methods seem consistent. Despite the shift towards the null, results suggest that the intervention improves objective response versus the control, statistically significantly at the 5\% level. In this case, the DR point estimate is not meaningfully different than the MAIC or G-computation point estimates, and the DR approach offers slightly increased standard errors and wider CIs than G-computation. Nevertheless, this loss of precision seems a relatively minor price to pay, compensated for by greater reassurance in our results due to increased protection against misspecification of the outcome model.  

\section{Discussion}\label{sec6}

The objective of this article was to clarify existing approaches for doubly robust estimation in the context of externally controlled SATs, and to propose an estimator that augments approaches based on MAIC or entropy balancing via an outcome model. We described and illustrated an extension of this estimator specifically tailored to unanchored ITCs, for the setting with unavailable external control IPD, which is commonly encountered in practice. In a simulation study and applied example, we evaluated the performance and demonstrated the use of different doubly robust augmented estimators, highlighting their merits with respect to the more popular non-augmented singly robust estimators. 

Our findings reinforce the understanding that ``balancing'' approaches to weighting, such as MAIC or entropy balancing, can enhance performance relative to standard ``modeling'' approaches, but lack the doubly robust property for non-linear outcome models. Conversely, augmented MAIC demonstrates double robustness and exhibits higher precision than non-augmented weighting estimators when the outcome model is correctly specified. Moreover, it generally achieves near-identical precision to G-computation, which offers the lowest variance under correct specification of the outcome model, but may exhibit notable bias where the outcome model is misspecified. 

A potential concern about augmented estimators has been possible bias and variance amplification where both the propensity score and the outcome model are incorrectly specified. Nevertheless, such amplification under dual model misspecification was not observed in our simulation study, and one can argue that risks are mitigated by employing ``balancing'' instead of ``modeling'' weights (for reasons outlined in the introduction to Section \ref{subsec33}), but further theoretical work and simulation studies are required to fully support this claim. The results of our simulation study motivate the routine application of doubly robust augmented estimators, particularly those based on MAIC or entropy balancing weights, in practical settings. It is unfortunate to see that virtually all applications of MAIC do not consider augmentation via an outcome model, and that most practical uses of augmented estimators apply ``modeling'' instead of ``balancing'' weights.\cite{webster2021using}

The weighted G-computation estimator described in Section \ref{subsec36} exhibited double robustness for the ATC when a canonical logistic link function was defined for the outcome model, but not when the link function was non-canonical, which coincides with what has been suggested previously.\cite{gabriel2024inverse} Future simulation studies should consider non-binary outcomes, including survival outcomes with censoring, and other summary effect measures beyond the log-odds ratio. 

Our simulation study and applied example considered scenarios with relatively low sample sizes in the SAT and the external control, corresponding to typical settings in rare disease and late-stage hematological or solid tumor oncology, where the number of subjects enrolled in SATs can be one- or two-hundred, but may also consist of several hundreds. The number of external controls can be equally small; as such, our findings are potentially applicable where the target estimand is the ATT instead of the ATC, which requires weighting the external control as opposed to the SAT. Somewhat worryingly, all augmented and non-augmented weighting estimators displayed some small-sample bias in our simulation study under a total sample size of $n=200$, even if modeling assumptions were correct.  This is consistent with previous research that has also noted small-sample bias.\cite{zetterqvist2019doubly, robins2020double}

The level of (deterministic) overlap between the SAT and external control covariate distributions in our simulation study was relatively strong. The performance of augmented and non-augmented weighting estimators with respect to G-computation will likely worsen as overlap decreases, particularly in conjunction with small sample sizes. Nevertheless, we conjecture that the performance of  balancing-based approaches will suffer less than that of their corresponding modeling-based counterparts, due to generating more stable and less extreme weights. This is unless a complete lack of overlap results in the absence of a solution to the convex optimization balancing problem, in which case extrapolating via G-computation might be the only option.

Finally, it is important to emphasize that all of the covariate-adjusted estimators we considered, including the doubly robust augmented approaches, require the important and untestable assumption of no unmeasured prognostic factors.  This is typically the main credibility concern of externally controlled SATs. In practice, important prognostic factors may be unknown or unavailable in at least one of the SAT or external control data sources. Researchers should always consider expert knowledge (e.g., consult clinicians, epidemiologists or statisticians with relevant expertise) and review the relevant literature to evaluate the plausibility of the no unmeasured prognostic factors assumption. An important area of future research is the development of sensitivity analysis or quantitative bias analysis methods to help explore the sensitivity of results to unmeasured prognostic factors, in the specific context of externally controlled SATs and unanchored ITCs.\cite{gupta2025quantitative}

\section*{Acknowledgments}

Not available. 

\subsection*{Financial disclosure}

No funding to report. 

\subsection*{AI disclosure}

During the preparation of this work, the authors used AI tools (Claude) for editing purposes.  The authors have reviewed and edited the final content.

\subsection*{Conflict of interest}

Harlan Campbell is employed by Precision AQ, a life sciences consultancy company, and Antonio Remiro-Az\'ocar is employed by Novo Nordisk, a pharmaceutical company. No conflicts of interest are declared as this research is purely methodological. 

\subsection*{Data Availability Statement}

\textcolor{black}{The files required to generate the data, run the simulations, and reproduce the results of the simulation study are available at \url{https://github.com/harlanhappydog/DRAWE-}. R code to reproduce the applied example is provided in the Supplementary Material.} 


\normalsize
\bibliography{wileyNJD-AMA}

\section{Appendix}

\subsection*{Consistency of weighting estimators}

To see that the IOW estimators are consistent under a correctly specified propensity score model for data source assignment, consider a simple scenario with a binary outcome, $Y$, and a single discrete covariate, $X$, such that:
\begin{align}
   \mu_{0}^{1} = \textrm{E}(Y|T=1,S=0) &= 1 \times \textrm{Pr}(Y=1|T=1,S=0) + {0 \times \textrm{Pr}(Y=0|T=1,S=0)} \nonumber \\
   &= \textrm{Pr}(Y=1|T=1,S=0). \nonumber\\
   \textrm{Basic probability rules imply that the marginal } & \textrm{risk is the weighted average of the stratum-specific risks:} \nonumber\\
   &= \sum_{x}\textrm{Pr}(Y=1|T=1,S=0, X=x) \textrm{Pr}(X=x|S=0, T=1). \nonumber\\
\textrm{Then, due to the assumption of conditional data}& \textrm{ source  ignorability, we have:} \nonumber\\ 
   &= \sum_{x}\textrm{Pr}(Y=1|T=1, X=x) \textrm{Pr}(X=x|S=0)\nonumber\\
      &= \sum_{x}\textrm{Pr}(Y=1|T=1,X=x) \textrm{Pr}(X=x|S=0) \frac{\textrm{Pr}(X=x|S=1)}{\textrm{Pr}(X=x|S=1)}\nonumber\\
       &= \sum_{x}\textrm{Pr}(Y=1|T=1,X=x) \textrm{Pr}(X=x|S=1) \frac{\textrm{Pr}(X=x|S=0)}{\textrm{Pr}(X=x|S=1)},\nonumber\\
\textrm{and from Bayes' rule, we have: } \quad \quad & \nonumber\\
       &=\sum_{x}\textrm{Pr}(Y=1|T=1,X=x) \textrm{Pr}(X=x|S=1) \frac{\textrm{Pr}(S=0|X=x)}{\textrm{Pr}(S=1|X=x)}\frac{\textrm{Pr}(S=1)}{\textrm{Pr}(S=0)} \nonumber\\
         &=\underbrace{\frac{\textrm{Pr}(S=1)}{\textrm{Pr}(S=0)}}_{A}\underbrace{\sum_{x}\textrm{Pr}(Y=1|T=1,X=x) \textrm{Pr}(X=x|S=1) \frac{\textrm{Pr}(S=0|X=x)}{\textrm{Pr}(S=1|X=x)}}_{B}.
             \end{align}

Since $\sum_{i=1}^{n}{\frac{S_{i}}{n}} = n_{1}/n \rightarrow \textrm{Pr}(S=1)$ and $\sum_{i=1}^{n} {(1-S_{i})}/n = n_{0}/n \rightarrow \textrm{Pr}(S=0)$, we can consistently estimate $A \approx {n_1}/{n_0}$. We can also consistently estimate $B$ from the sample since the covariate distribution for the SAT, $(X|S=1)$, is observed (and since, within the sample, $T=1 \iff S=1$):
\begin{align*}
B&\approx \sum_{i=1}^{n_{1}}\textrm{Pr}(Y=1|S=1,X=x_{i}) \frac{\textrm{Pr}(S=0|X=x_{i})}{\textrm{Pr}(S=1|X=x_{i})}\approx\sum_{i=1}^{n_{1}}\frac{Y_{i}}{n_{1}} \frac{\textrm{Pr}(S=0|X=x_{i})}{\textrm{Pr}(S=1|X=x_{i})}.\nonumber
         \end{align*}
Finally,  if the propensity score model is correctly specified, the inverse odds weights consistently estimate the true inverse odds, such that, for $i$ in 1,...,$n_{1}$:
                \begin{align}
 \widehat{w}_{i} \rightarrow& \frac{\textrm{Pr}(S=0|X=x_{i})}{\textrm{Pr}(S=1|X=x_{i})} .
  \end{align}
  Therefore, the IOW estimator is consistent: 
     \begin{align}
\hat{\mu}_{0}^{1} &= \underbrace{\frac{n_{1}}{n_{0}}}_{A} \underbrace{\sum_{i=1}^{n_{1}}{\frac{Y_{i}}{n_{1}}\hat{w}_{i}} }_{B}\nonumber \\
&= \frac{1}{n_{0}}\sum_{i=1}^{n_{1}}{Y_{i} \hat{w}_{i}} \rightarrow \mu_{0}^{1}.
         \end{align}
Note that $\textrm{E}(S_{i}w_{i}) =  n_{0}/n_{1}$ implies that $\textrm{E}\left(\frac{1}{n_{1}}\sum_{i=1}^{n_{1}}{w_{i}}\right) =  n_{0}/n_{1}$, which implies that $\textrm{E}\left(\sum_{i=1}^{n_{1}}{w_{i}}\right) =  n_{0}$. As such, the normalized IOW estimator in Equation \ref{norm_weights_2} is also consistent for the ATC.

While the entropy balancing MAIC weights ($\hat{v}_{i}$) defined in Equation \ref{weight_equation} will be different than the IOW weights ($\widehat{w}_{i}$) obtained from maximum-likelihood estimation of the logistic regression model in Equation \ref{logistic_PS}, $\hat{\nu}_{i} = n_{0}\hat{v}_{i}$ will consistently estimate the true inverse odds if the logistic regression model is correctly specified.  To be clear, if the logistic regression model correctly specifies the true propensity score model, then we have both: $\widehat{w}_i \rightarrow \frac{\textrm{Pr}(S=0|X=x_{i})}{\textrm{Pr}(S=1|X=x_{i})}$ (i.e., the IOW weights are consistent) and  $\widehat{\nu}_i \rightarrow \frac{\textrm{Pr}(S=0|X=x_{i})}{\textrm{Pr}(S=1|X=x_{i})}$ (i.e., the entropy balancing weights are consistent), for $i$ in 1,...,$n_{1}$; see Zhao and Percival (2017) \cite{zhao2017entropy} for details.  Therefore, following the same logic as detailed in Section \ref{subsec32} for the IOW estimators, the MAIC estimator, as defined in Equation \ref{norm_weights_3}, is also consistent if the implied propensity score model is correctly specified.

\subsection*{Double robustness of the augmented weighting estimators}

Note that the following derivation does not depend on using the canonical link function for the outcome model. Consider the simple scenario where we have a binary outcome, $Y$, and a single discrete covariate, $X$, and the augmented weighting estimator in Equation \ref{absolute_maicdr} is re-written as:
\begin{equation}
\hat{\mu}_0^1
=
\sum_{i=1}^{n_1}
\hat{v}_{i}
\left
(
Y_i
-
\hat{Y}^1_i
\right )
+
\frac{1}{n_0}
\sum_{i=n_1+1}^{n}
\hat{Y_i}^1.
\label{general_form_app}
\end{equation}
If the outcome model is correctly specified, the expectation of the first summation in Equation \ref{general_form_app} converges to zero as $n_1 \rightarrow \infty$ because $\hat{Y}_i^1 \rightarrow Y_{i}$ and the terms inside the summation cancel out, irrespective of any postulated propensity score model. The second summation is equivalent to the G-computation estimator and is consistent for $\mu_0^1$ because the outcome model is correct. Consequently, $\hat{\mu}_0^1 \rightarrow \mu_0^1$, and $\widehat{\textrm{ATC}} \rightarrow \textrm{ATC}$  (assuming $\hat{\mu}_0^0 \rightarrow \mu_0^0$).

Conversely, if the propensity score model is correctly specified but the outcome model is incorrect, the first summation consistently cancels out the bias produced by the G-computation estimator in the second summation and the remainder term is exactly equal to the non-augmented weighting estimator, which converges to $\mu_0^1$ as $n_1 \rightarrow \infty$ because the propensity score model is correct.  To illustrate this, consider rearranging Equation \ref{general_form_app} to: 

\begin{align}
\hat{\mu}_0^1
&=
\sum_{i=1}^{n_1}
\hat{v}_{i}
Y_i
-
\sum_{i=1}^{n_1}
\hat{v}_{i}
\hat{Y}^1_i
+
\frac{1}{n_0}
\sum_{i=n_1+1}^{n}
\hat{Y_i}^1 \nonumber \\
&=
\underbrace{
\sum_{i=1}^{n_1}
\hat{v}_{i}
Y_i}_C
+
\underbrace{
\left (
\frac{1}{n_0}
\sum_{i=n_1+1}^{n}
\hat{Y_i}^1 
-
\sum_{i=1}^{n_1}
\hat{v}_{i}
\hat{Y}^1_i
\right )}_D. 
\label{general_form_2}
\end{align}

First, $C$ is equivalent to the MAIC estimator and is consistent for $\mu_0^1$ because the propensity score model is correct.  Expanding the summations in $D$ over $i=1,...,n$, we have:
\begin{align}
D &=  \left(\sum_{i=1}^{n}
\frac{(1-S_{i})\hat{Y}_{i}^{1}}{n_{0}} -\frac{S_{i}\hat{\nu}_{i}\hat{Y}_{i}^{1}}{n_{0}} \right)\\
&= \frac{1}{n_{0}}\left(\sum_{i=1}^{n}{\hat{Y}_{i}^{1}}
\left ((1-S_{i})-S_{i}\hat{\nu}_{i} 
\right )\right),
\end{align}
where $\hat{\nu}_{i} = n_{0}\hat{v}_{i}$, for $i$ in 1,...,$n$. Then, suppose that the outcome model for $Y^1$ is independent of $S$ and that $Y^{1*}_{i}$ is the large sample limit of $\hat{Y}^{1}_{i}$. Also, suppose that the large sample limit of $\hat{\nu}_{i}$ is $\frac{\textrm{Pr}(S=0|X=x_{i})}{\textrm{Pr}(S=1|X=x_{i})}$, since the propensity score model is correctly specified. Then: 
\begin{align}
\textrm{E}\left(\hat{Y}_{i}^{1}
\left (
(1-S_{i})-S_{i}\hat{\nu}_{i}
\right )\right ) &= 
\textrm{E}
\left (
\textrm{E}
\left
(\hat{Y}_{i}^{1}
\left (
(1-S_{i})-S_{i}\hat{\nu}_{i}
\right 
)\mid X
\right )
\right ) \nonumber \\
& =\textrm{E}
\left(\textrm{E}
(\hat{Y}_{i}^{1} \mid X)
\times
\textrm{E}
\left(
\left (
(1-S_{i})-S_{i}\hat{\nu}_{i}
\right )\mid X \right )\right ) \nonumber \\
& \rightarrow \textrm{E}
\left ( Y^{1*}_{i}\sum_{x}{\textrm{Pr}(S=0|X=x)\textrm{Pr}(X=x) - \textrm{Pr}(S=1|X=x)\textrm{Pr}(X=x)\frac{\textrm{Pr}(S=0|X=x)}{\textrm{Pr}(S=1|X=x)}}\right ) \nonumber\\
&=\textrm{E}\left (Y^{1*}_{i}\sum_{x}{\textrm{Pr}(S=0|X=x)\textrm{Pr}(X=x) - \textrm{Pr}(X=x)\textrm{Pr}(S=0|X=x)}\right ) \nonumber \\ 
&=\textrm{E}\left(Y^{1*}_{i} \times 0\right) \nonumber\\ 
&=0.
\end{align}
Consequently, $D\rightarrow0$ which implies that $\hat{\mu}_0^1 \rightarrow \mu_0^1$, and $\widehat{\textrm{ATC}} \rightarrow \textrm{ATC}$ having assumed $\hat{\mu}_0^0 \rightarrow \mu_0^0$.

\clearpage
\section*{Supplementary Material}

\subsection*{Targeting the average treatment effect in the treated}

We briefly adapt the methodologies in Section \ref{subsec32} to Section \ref{subsec36} of the main text so that these target the ATT. We assume that there is full IPD availability and that $\hat{\mu}_1^1=\frac{1}{n_1}\sum_{i=1}^{n_1} Y_i$ is consistent for $\mu_1^1$. 

For the modeling-based IOW approaches in Section \ref{subsec32}, external control subjects are weighted by their conditional odds of SAT participation to transport the external control outcomes to the SAT (sub) population. SAT subjects are unweighted and external control subjects $i=n_1+1,\dots, n$ are weighted by $\hat{w}_i=\hat{e}_i/(1-\hat{e}_i)$. Assuming correct specification of the propensity score model, the estimated weights would balance the covariate distribution of the external control with respect to that of the SAT, enabling consistent estimation of mean absolute outcome $\mu^0_1$ and the ATT. Propensity score predictions that are close to one lead to extreme weights and imprecise ATT estimation, particularly where the sample size of the external control is small. 

A MAIC estimator for the ATT, akin to that described in Section \ref{subsec33}, would enforce that the covariate distributional features of the weighted external control subjects are exactly balanced with respect to those of the SAT subjects. As such, the balancing constraints would center the external control covariate balance functions on their SAT means. MAIC enables consistent estimation of $\mu^0_1$ and the ATT, as long as either the log-odds of the propensity score or the potential outcome under the control are linear across the specified balance functions. The general form of the weighting estimators for the ATT is:
\begin{equation*}
\widehat{\textrm{ATT}} 
= 
g 
\underbrace
{
\left ( \frac{1}{n_1} \sum_{i=1}^{n_1} Y_i
\right )
}
_{\hat{\mu}_1^1}
-
g
\underbrace
{
\left (
\frac{1}{K}\sum_{i=n_1+1}^n \hat{v}_iY_i
\right )
}
_{\hat{\mu}_1^0}
,
\end{equation*}
where $K$ is a constant and $\hat{v}_i$ is a weight estimate  for $i=n+1,\dots,n$, derived using the modeling approach or MAIC. 

A G-computation estimator such like that described in Section \ref{subsec34} but for the ATT requires postulating a model for the potential outcome expectation under the control, fitted to the external control participants. Based on the fitted model $m({\mathbf{X}_i};\hat{\mathbf{\beta}})$, the potential outcome under the control is predicted for each subject $i=1,\dots, n_1$ in the SAT: $\hat{Y}_i^0 = q^{-1} \left (m({\mathbf{X}_i};\hat{\mathbf{\beta}})\right)$. The potential outcome predictions are averaged over the empirical covariate distribution of the SAT, resulting in the ATT estimator: 
\begin{equation}
\widehat{\textrm{ATT}} 
= 
g 
\underbrace
{
\left ( \frac{1}{n_1} \sum_{i=1}^{n_1} Y_i
\right )
}
_{\hat{\mu}_1^1}
-
g 
\underbrace
{
\left ( \frac{1}{n_1} \sum_{i=1}^{n_1} \hat{Y}^0_i
\right )
}
_{\hat{\mu}_1^0},
\label{GComp_ATT}
\end{equation}
which relies on correct specification of the model for the potential outcome under the control for consistent estimation. 

The DR augmented weighting estimators, proposed in Section \ref{subsec35}, would target the ATT as follows. Based on an outcome model $m({\mathbf{X}_i};\hat{\mathbf{\beta}})$ fitted to the external control participants, the potential outcome under the control is predicted for all subjects $i=1,\dots,n$·in the SAT and the external control: $\hat{Y}_i^0 = q^{-1} \left (m({\mathbf{X}_i};\hat{\mathbf{\beta}})\right)$. The potential outcome predictions are augmented with a weighted average of residuals for the external control subjects. The general form of the doubly robust augmented weighting estimators for the ATT is:
\begin{equation*}
\widehat{\textrm{ATT}} = 
g 
\underbrace
{
\left (
\frac{1}{n_1}\sum_{i=1}^{n_1} Y_i
\right )
}_{\hat{\mu}_1^1}
-
g 
\underbrace
{
\left (
\frac{1}{K}
\sum_{i=n_1+1}^{n}
\hat{u}
_i
\epsilon_i^0
+
\frac{1}{n_1}
\sum_{i=1}^{n_1}
\hat{Y_i}^0
\right )
}
_{\hat{\mu}_1^0},
\end{equation*}
where $K$ is a constant, $\hat{u}_i$ is a weight estimate and $\epsilon_i^0=Y_i -\hat{Y}_i^0$ is a residual term for subject $i=n_1+1,\dots,n$ in the external control. 

A weighted G-computation estimator akin to that described in Section \ref{subsec36} would target the ATT by: (1) estimating weights for the odds of SAT participation; (2) fitting a weighted model $m({\mathbf{X}_i};\hat{\mathbf{\beta}}_v)$ for the conditional outcome expectation to the external control participants; and (3) averaging the outcome predictions of the weighted regression over the SAT covariate distribution. The resulting estimator for the mean absolute outcome $\mu_1^0$ is $\hat{\mu}_1^0= 
\frac{1}
{n_1}
\sum_{i=1}^{n_1}
\hat{Y}_i^0
=
\frac{1}
{n_1}
\sum_{i=1}^{n_1}
q^{-1} \left (m({\mathbf{X}_i};\hat{\mathbf{\beta}}_v)\right)$, which is then substituted into Equation \ref{GComp_ATT} for estimation of the ATT.

\subsection*{Plots of covariate overlap for the simulation study}

\begin{figure}[h]
    \centering
    \includegraphics[width=0.5\linewidth]{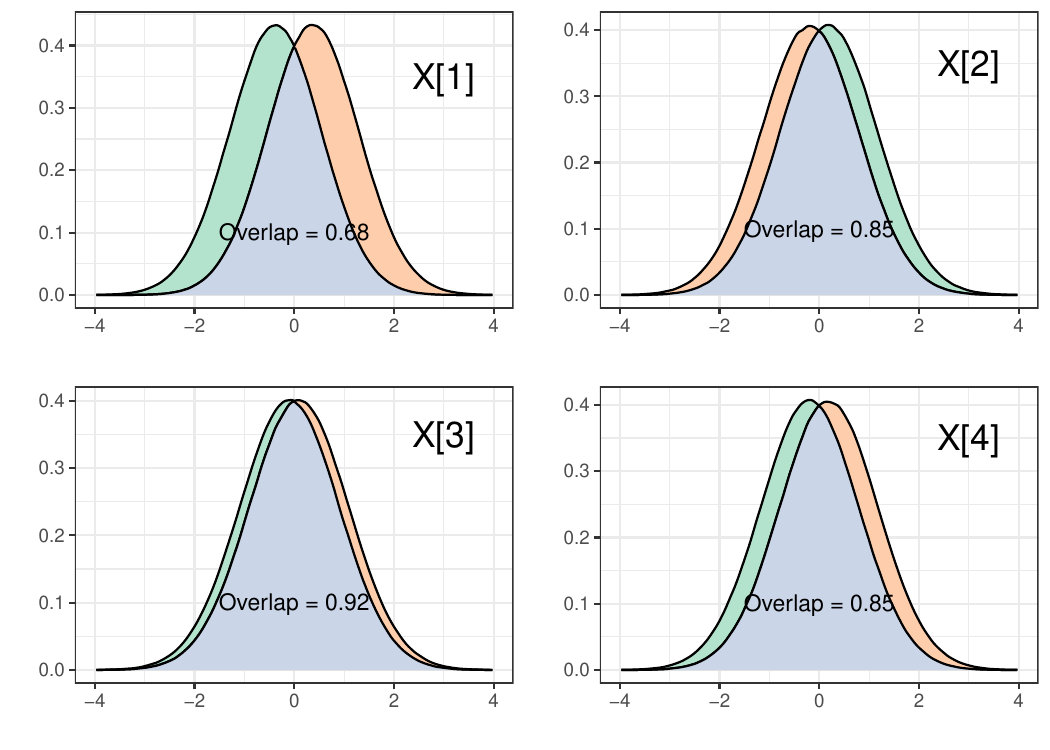}
    \caption{Density plots showing the overlap of covariates $X_{1}$,$X_{2}$,$X_{3}$, and $X_{4}$ for Scenarios KS1 and KS2 in the simulation study. }
    \label{fig:overlap1}
\end{figure}

\begin{figure}[!htb]
    \centering
    \includegraphics[width=0.5\linewidth]{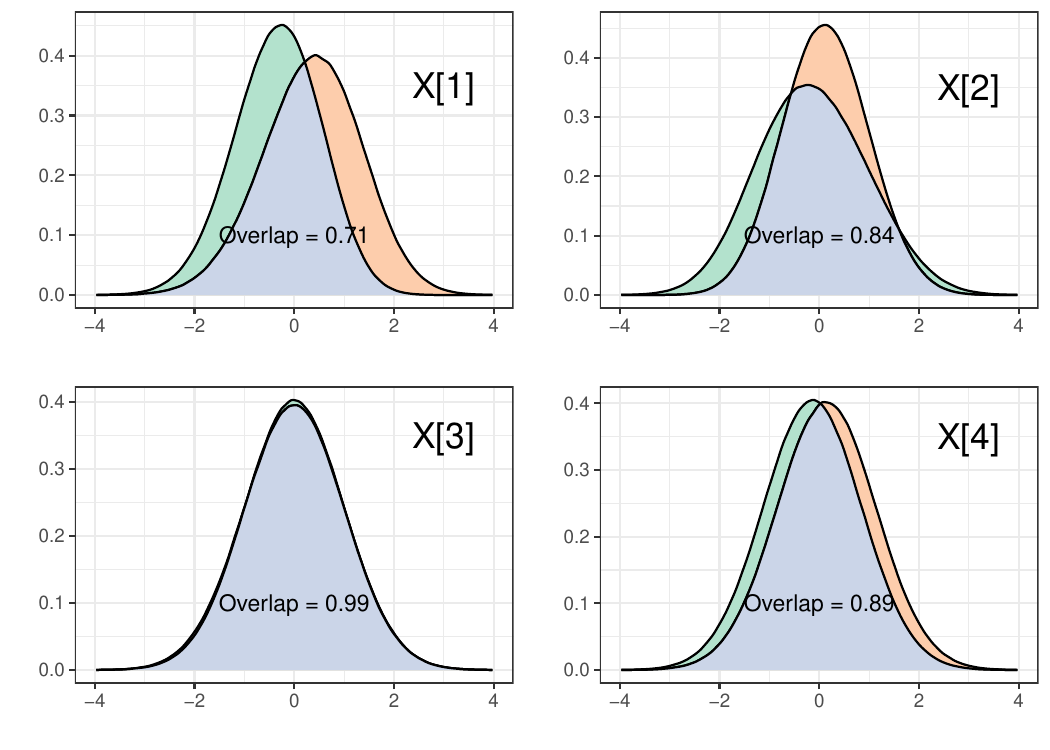}
    \caption{Density plots showing the overlap of covariates $X_{1}$,$X_{2}$,$X_{3}$, and $X_{4}$ for Scenarios KS3 and KS4 in the simulation study. }
    \label{fig:overlap3}
\end{figure}

\subsection*{R code for the applied example}

The following R code provides all calculations in the applied example:  
\begin{footnotesize}
\begin{verbatim}
library(dplyr)
library(boot)
library(MAIC)
library(ggplot2)
library(maicChecks)

set.seed(1894)
boot_n <- 10000

# g-function is the log-odds:
g_function <- function(p){log(p/(1-p))}

#### Intervention data

adsl <- read.csv(system.file("extdata", "adsl.csv", package = "MAIC", mustWork = TRUE))
adrs <- read.csv(system.file("extdata", "adrs.csv", package = "MAIC", mustWork = TRUE))

adsl <- adsl %>%
  mutate(SEX = ifelse(SEX == "Male", 1, 0))

adrs <- adrs %>%
  filter(PARAM == "Response") %>%
  transmute(USUBJID, ARM, response = AVAL)

intervention_input <- adsl %>%
  full_join(adrs, by = c("USUBJID", "ARM"))

# Baseline aggregate data for the comparator population
target_pop <- read.csv(system.file("extdata", "aggregate_data.csv",
                                   package = "MAIC", mustWork = TRUE))

# Simulate response data based on the known proportion of responders
comparator_n <- target_pop$N
comparator_prop_events <- 0.4
n_with_event <- round(comparator_n * comparator_prop_events, digits = 0)

n0 <- comparator_n
n1 <- nrow(intervention_input)
Y_all <- c(intervention_input$response, rep(1, n_with_event), rep(0, n0 - n_with_event))
S_all <- c(rep(1, n1), rep(0, n0))
X_all <- rbind(
  intervention_input %>% select(AGE, SEX, SMOKE, ECOG0),
  data.frame(AGE = rep(NA, n0), SEX = rep(NA, n0),
             SMOKE = rep(NA, n0), ECOG0 = rep(NA, n0))
)
X_all$AGE_SQ <- (X_all$AGE)^2

#############################################
# Naive estimate
#############################################

mu1_naive <- mean(Y_all[S_all == 1])
mu0_naive <- mean(Y_all[S_all == 0])
ATC_naive <- g_function(mu1_naive) - g_function(mu0_naive)

SE_g_mu1 <- sqrt(1 / (n1 * mu1_naive * (1 - mu1_naive)))
SE_g_mu0 <- sqrt(1 / (n0 * mu0_naive * (1 - mu0_naive)))
SE_ATC_naive <- sqrt(SE_g_mu1^2 + SE_g_mu0^2)
ATC_naive_CI <- c(ATC_naive - qnorm(0.975) * SE_ATC_naive,
                  ATC_naive + qnorm(0.975) * SE_ATC_naive)

round(c(ATC_naive, ATC_naive_CI), 3)
round(exp(c(ATC_naive, ATC_naive_CI)), 3)

#############################################
# Entropy balancing (MAIC)
#############################################

objfn <- function(a1, X){ sum(exp(X %*% a1)) }
gradfn <- function(a1, X){ colSums(sweep(X, 1, exp(X %*% a1), "*")) }
cov_names <- paste0("X.", colnames(X_all))
AC.IPD <- data.frame(y = Y_all[S_all == 1], X = X_all[S_all == 1, ])

BC.ALD <- data.frame(cbind(target_pop$age.mean,
                           target_pop$prop.male,
                           target_pop$prop.smoke,
                           target_pop$prop.ecog0,
                           target_pop$age.mean^2 + target_pop$age.sd^2))
colnames(BC.ALD) <- paste0("mean.", cov_names)

maicLP(AC.IPD[, -1], BC.ALD)

X.EM.0 <- sweep(as.matrix(AC.IPD[, cov_names]), 2,
                as.matrix(BC.ALD[, paste0("mean.", cov_names)]), '-')

gamma <- optim(par = rep(0, ncol(X.EM.0)),
               fn = objfn, gr = gradfn, X = X.EM.0, method = "BFGS")$par
wt_EB <- exp(X.EM.0 %*% gamma) / sum(exp(X.EM.0 %*% gamma))

mu1_EB <- sum(wt_EB * Y_all[S_all == 1])
ATC_EB <- g_function(mu1_EB) - g_function(mu0_naive)

ATC_EB_function <- function(data, indices){
  temp <- data[indices, ]
  AC.IPD <- data.frame(y = temp[, "Y_all"][temp[, "S_all"] == 1],
                       X = temp[temp[, "S_all"] == 1, grep("X_all", colnames(temp))])
  BC.ALD <- data.frame(cbind(target_pop$age.mean,
                             target_pop$prop.male,
                             target_pop$prop.smoke,
                             target_pop$prop.ecog0,
                             target_pop$age.mean^2 + target_pop$age.sd^2))
  objfn <- function(a1, X){ sum(exp(X %*% a1)) }
  gradfn <- function(a1, X){ colSums(sweep(X, 1, exp(X %*% a1), "*")) }
  cov_names <- paste0("X.X_all.", colnames(X_all))
  colnames(BC.ALD) <- paste0("mean.", cov_names)
  X.EM.0 <- sweep(as.matrix(AC.IPD[, cov_names]), 2,
                  as.matrix(BC.ALD[, paste0("mean.", cov_names)]), '-')
  gamma <- optim(par = rep(0, ncol(X.EM.0)),
                 fn = objfn, gr = gradfn, X = X.EM.0, method = "BFGS")$par
  wt_EB <- exp(X.EM.0 %*% gamma) / sum(exp(X.EM.0 %*% gamma))
  mu1_EB <- sum(wt_EB * temp[, "Y_all"][temp[, "S_all"] == 1])
  return(g_function(mu1_EB))
}

set.seed(123)
boot_samples <- boot(data = data.frame(Y_all = Y_all, S_all = S_all, X_all = X_all),
                     statistic = ATC_EB_function, R = boot_n,
                     strata = S_all, parallel = "multicore")
SE_ATC_EB <- sqrt(sd(boot_samples$t, na.rm = TRUE)^2 + SE_g_mu0^2)
ATC_EB_CI <- c(ATC_EB - qnorm(0.975) * SE_ATC_EB,
               ATC_EB + qnorm(0.975) * SE_ATC_EB)

round(c(ATC_EB, ATC_EB_CI), 3)
round(exp(c(ATC_EB, ATC_EB_CI)), 3)

#############################################
# Simulate M individual values from target population
#############################################

M <- 10000
set.seed(123)
out2 <- add_integration(
  data.frame(Y_all = NA),
  AGE   = distr(qnorm, mean = target_pop$age.mean, sd = target_pop$age.sd),
  SEX   = distr(qbern, prob = target_pop$prop.male),
  SMOKE = distr(qbern, prob = target_pop$prop.smoke),
  ECOG0 = distr(qbern, prob = target_pop$prop.ecog0),
  cor   = cor(X_all[S_all == 1, c("AGE", "SEX", "SMOKE", "ECOG0")]),
  n_int = M
)

x_star <- cbind(unlist(out2$.int_AGE),
                unlist(out2$.int_SEX),
                unlist(out2$.int_SMOKE),
                unlist(out2$.int_ECOG0))
x_star <- cbind(x_star, (x_star[, 1])^2)
colnames(x_star) <- colnames(X_all[S_all == 1, ])

n_with_event <- round(M * comparator_prop_events, digits = 0)
Y_all <- c(Y_all[S_all == 1], rep(1, n_with_event), rep(0, M - n_with_event))
X_all <- rbind(X_all[S_all == 1, ], x_star)
S_all <- c(S_all[S_all == 1], rep(0, M))

n1 <- sum(S_all == 1)
n0 <- sum(S_all == 0)

#############################################
# G-computation estimator
#############################################

outcome_model <- glm(y ~ .,
                     data = data.frame(y = Y_all[S_all == 1],
                                       x = X_all[S_all == 1, ]),
                     family = binomial(link = "logit"))

Y1_hat <- predict(outcome_model, newdata = data.frame(x = X_all[S_all == 0, ]),
                  type = "response")

mu1_GCOMP <- (1 / n0) * sum(Y1_hat)
ATC_GCOMP <- g_function(mu1_GCOMP) - g_function(mu0_naive)

ATC_GCOMP_function <- function(data, indices){
  temp <- data[indices, ]
  outcome_model <- glm(y ~ .,
                       data = data.frame(y = temp[, "Y_all"][temp[, "S_all"] == 1],
                                         x = temp[temp[, "S_all"] == 1, grep("X_all", colnames(temp))]),
                       family = "binomial")
  Y1_hat <- predict(outcome_model,
                    newdata = data.frame(x = temp[temp[, "S_all"] == 0, grep("X_all", colnames(temp))]),
                    type = "response")
  mu1_GCOMP <- (1 / sum(temp[, "S_all"] == 0)) * sum(Y1_hat)
  return(g_function(mu1_GCOMP))
}

set.seed(123)
boot_samples <- boot(data = data.frame(Y_all = Y_all, S_all = S_all, X_all = X_all),
                     statistic = ATC_GCOMP_function, R = boot_n,
                     strata = S_all, parallel = "multicore")
SE_ATC_GCOMP <- sqrt(sd(boot_samples$t, na.rm = TRUE)^2 + SE_g_mu0^2)
ATC_GCOMP_CI <- c(ATC_GCOMP - qnorm(0.975) * SE_ATC_GCOMP,
                  ATC_GCOMP + qnorm(0.975) * SE_ATC_GCOMP)

round(c(ATC_GCOMP, ATC_GCOMP_CI), 3)
round(exp(c(ATC_GCOMP, ATC_GCOMP_CI)), 3)

#############################################
# DR augmented MAIC estimator
#############################################

data_for_outcome_model <- data.frame(y = Y_all[S_all == 1], X_all[S_all == 1, ])
colnames(data_for_outcome_model) <- c("y", colnames(X_all))
outcome_model <- glm(y ~ ., data = data_for_outcome_model, family = "binomial")
Y1_hat_all <- predict(outcome_model, newdata = data.frame(X_all), type = "response")

mu1_DR3 <- (1 / sum(wt_EB)) * sum(wt_EB * (Y_all[S_all == 1] - Y1_hat_all[S_all == 1])) +
  (1 / n0) * sum(Y1_hat_all[S_all == 0])
ATC_DR3 <- g_function(mu1_DR3) - g_function(mu0_naive)

ATC_DR3_function <- function(data, indices){
  temp <- data[indices, ]
  outcome_model <- glm(y ~ .,
                       data = data.frame(y = temp[, "Y_all"][temp[, "S_all"] == 1],
                                         x = temp[temp[, "S_all"] == 1, grep("X_all", colnames(temp))]),
                       family = "binomial")
  Y1_hat_all <- predict(outcome_model,
                        newdata = data.frame(x = temp[, grep("X_all", colnames(temp))]),
                        type = "response")
  AC.IPD <- data.frame(y = temp[, "Y_all"][temp[, "S_all"] == 1],
                       X = temp[temp[, "S_all"] == 1, grep("X_all", colnames(temp))])
  BC.ALD <- data.frame(matrix(apply(temp[temp[, "S_all"] == 0, grep("X_all", colnames(temp))], 2, mean), 1, ))
  objfn <- function(a1, X){ sum(exp(X %*% a1)) }
  gradfn <- function(a1, X){ colSums(sweep(X, 1, exp(X %*% a1), "*")) }
  cov_names <- paste0("X.X_all.", colnames(X_all))
  colnames(BC.ALD) <- paste0("mean.", cov_names)
  X.EM.0 <- sweep(as.matrix(AC.IPD[, cov_names]), 2,
                  as.matrix(BC.ALD[, paste0("mean.", cov_names)]), '-')
  gamma <- optim(par = rep(0, ncol(X.EM.0)),
                 fn = objfn, gr = gradfn, X = X.EM.0, method = "BFGS")$par
  wt_EB <- exp(X.EM.0 %*% gamma) / sum(exp(X.EM.0 %*% gamma))
  mu1_DR3 <- (1 / sum(wt_EB)) * sum(wt_EB * (temp[, "Y_all"][temp[, "S_all"] == 1] -
                                               Y1_hat_all[temp[, "S_all"] == 1])) +
    (1 / sum(temp[, "S_all"] == 0)) * sum(Y1_hat_all[temp[, "S_all"] == 0])
  return(g_function(mu1_DR3))
}

set.seed(123)
boot_samples <- boot(data = data.frame(Y_all = Y_all, S_all = S_all, X_all = X_all),
                     statistic = ATC_DR3_function, R = boot_n,
                     strata = S_all, parallel = "multicore")
SE_ATC_DR3 <- sqrt(sd(boot_samples$t, na.rm = TRUE)^2 + SE_g_mu0^2)
ATC_DR3_CI <- c(ATC_DR3 - qnorm(0.975) * SE_ATC_DR3,
                ATC_DR3 + qnorm(0.975) * SE_ATC_DR3)

round(c(ATC_DR3, ATC_DR3_CI), 3)
round(exp(c(ATC_DR3, ATC_DR3_CI)), 3)

#############################################
# Normalized Inverse Odds Weighting (Hajek type)
#############################################

ps_model <- glm(S ~ AGE + SEX + SMOKE + ECOG0 + AGE_SQ,
                data = data.frame(S = S_all, X_all),
                family = binomial(link = "logit"))

e_hat_intervention <- predict(ps_model, type = "response")[S_all == 1]
iow_raw <- (1 - e_hat_intervention) / e_hat_intervention
iow_normalized <- iow_raw / sum(iow_raw)

# Effective sample size
ESS_IOW <- (sum(iow_raw))^2 / sum(iow_raw^2)

# ATC estimation
mu1_IOW_Hajek <- sum(iow_normalized * intervention_input$response)
ATC_IOW_Hajek <- g_function(mu1_IOW_Hajek) - g_function(mu0_naive)

# Bootstrap for SE
ATC_IOW_Hajek_function <- function(data, indices) {
  temp <- data[indices, ]
  ps_model_b <- glm(S ~ AGE + SEX + SMOKE + ECOG0 + AGE_SQ,
                    data = temp, family = binomial(link = "logit"))
  e_hat_b <- predict(ps_model_b, type = "response")
  e_hat_int_b <- e_hat_b[temp$S == 1]
  iow_raw_b <- (1 - e_hat_int_b) / e_hat_int_b
  iow_norm_b <- iow_raw_b / sum(iow_raw_b)
  mu1_b <- sum(iow_norm_b * temp$Y[temp$S == 1])
  mu1_b <- max(min(mu1_b, 0.9999), 0.0001)
  return(g_function(mu1_b))
}

set.seed(123)
boot_samples_IOW <- boot(
  data = data.frame(Y = Y_all, S = S_all, X_all),
  statistic = ATC_IOW_Hajek_function,
  R = boot_n, strata = S_all, parallel = "multicore"
)
SE_ATC_IOW <- sqrt(sd(boot_samples_IOW$t, na.rm = TRUE)^2 + SE_g_mu0^2)
ATC_IOW_Hajek_CI <- c(ATC_IOW_Hajek - qnorm(0.975) * SE_ATC_IOW,
                      ATC_IOW_Hajek + qnorm(0.975) * SE_ATC_IOW)

round(c(ATC_IOW_Hajek, ATC_IOW_Hajek_CI), 3)
round(exp(c(ATC_IOW_Hajek, ATC_IOW_Hajek_CI)), 3)

#############################################
# Covariate balance table
#############################################

fmt2 <- function(x) sprintf("%.2f", x)

# MAIC-weighted statistics
ESS_MAIC <- 1 / sum(wt_EB^2)
maic_mean_age   <- sum(wt_EB * intervention_input$AGE)
maic_mean_sex   <- sum(wt_EB * intervention_input$SEX)
maic_mean_ecog  <- sum(wt_EB * intervention_input$ECOG0)
maic_mean_smoke <- sum(wt_EB * intervention_input$SMOKE)
maic_sd_age     <- sqrt(sum(wt_EB * (intervention_input$AGE - maic_mean_age)^2))

# IOW-weighted statistics
iow_mean_age   <- sum(iow_normalized * intervention_input$AGE)
iow_mean_sex   <- sum(iow_normalized * intervention_input$SEX)
iow_mean_ecog  <- sum(iow_normalized * intervention_input$ECOG0)
iow_mean_smoke <- sum(iow_normalized * intervention_input$SMOKE)
iow_sd_age     <- sqrt(sum(iow_normalized * (intervention_input$AGE - iow_mean_age)^2))

balance_df <- data.frame(
  Covariate = c("Age in years (mean; SD)", "Sex (proportion male)",
                "ECOG (proportion status 1)", "Smoking (proportion smokers)"),
  Intervention = c(
    paste0(fmt2(mean(intervention_input$AGE)), "; ", fmt2(sd(intervention_input$AGE))),
    fmt2(mean(intervention_input$SEX)),
    fmt2(mean(intervention_input$ECOG0)),
    fmt2(mean(intervention_input$SMOKE))),
  External = c(
    paste0(fmt2(target_pop$age.mean), "; ", fmt2(target_pop$age.sd)),
    fmt2(target_pop$prop.male), fmt2(target_pop$prop.ecog0), fmt2(target_pop$prop.smoke)),
  IOW = c(
    paste0(fmt2(iow_mean_age), "; ", fmt2(iow_sd_age)),
    fmt2(iow_mean_sex), fmt2(iow_mean_ecog), fmt2(iow_mean_smoke)),
  MAIC = c(
    paste0(fmt2(maic_mean_age), "; ", fmt2(maic_sd_age)),
    fmt2(maic_mean_sex), fmt2(maic_mean_ecog), fmt2(maic_mean_smoke)),
  stringsAsFactors = FALSE
)

print(balance_df)

#############################################
# Combined histogram of weights
#############################################

weights_df <- data.frame(
  Weight = c(as.numeric(wt_EB), as.numeric(iow_normalized)),
  Method = factor(
    c(rep("MAIC (entropy balancing)", length(wt_EB)),
      rep("Normalized IOW", length(iow_normalized))),
    levels = c("Normalized IOW", "MAIC (entropy balancing)")
  )
)

p_hist <- ggplot(weights_df, aes(x = Weight)) +
  geom_histogram(bins = 30, fill = "grey70", colour = "black", linewidth = 0.3) +
  facet_wrap(~ Method, scales = "free", ncol = 2) +
  labs(x = "Weight", y = "Frequency") +
  theme_minimal(base_size = 12) +
  theme(strip.text = element_text(face = "bold", size = 11),
        panel.grid.minor = element_blank())

print(p_hist)

#############################################
# Forest plot comparing all estimators
#############################################

forest_data <- data.frame(
  Estimator = factor(
    c("Naive", "IOW (Hajek)", "MAIC (EB)", "G-computation", "DR augmented MAIC"),
    levels = c("DR augmented MAIC", "G-computation", "MAIC (EB)", "IOW (Hajek)", "Naive")
  ),
  Estimate  = c(ATC_naive, ATC_IOW_Hajek, ATC_EB, ATC_GCOMP, ATC_DR3),
  CI_lower  = c(ATC_naive_CI[1], ATC_IOW_Hajek_CI[1], ATC_EB_CI[1], ATC_GCOMP_CI[1], ATC_DR3_CI[1]),
  CI_upper  = c(ATC_naive_CI[2], ATC_IOW_Hajek_CI[2], ATC_EB_CI[2], ATC_GCOMP_CI[2], ATC_DR3_CI[2])
)

forest_data$label <- paste0(
  sprintf("%.3f", forest_data$Estimate), " (",
  sprintf("%.3f", forest_data$CI_lower), ", ",
  sprintf("%.3f", forest_data$CI_upper), ")"
)

p_logOR <- ggplot(forest_data, aes(x = Estimate, y = Estimator)) +
  geom_vline(xintercept = 0, linetype = "dashed", colour = "grey50") +
  geom_errorbarh(aes(xmin = CI_lower, xmax = CI_upper),
                 height = 0.2, linewidth = 0.6) +
  geom_point(size = 4, shape = 20) +
  geom_text(aes(x = max(forest_data$CI_upper) + 0.15, label = label),
            hjust = 0, size = 3.2) +
  labs(x = "ATC (marginal log-odds ratio)", y = NULL) +
  theme_minimal(base_size = 12) +
  theme(panel.grid.major.y = element_blank(),
        panel.grid.minor = element_blank(),
        axis.text.y = element_text(size = 11)) +
  coord_cartesian(xlim = c(min(forest_data$CI_lower) - 0.1,
                           max(forest_data$CI_upper) + 1.2))

print(p_logOR)
  
\end{verbatim}
\end{footnotesize}

\end{document}